\documentclass[12pt]{article}
\usepackage{color}
\usepackage{amsmath}
\usepackage{amsfonts}
\usepackage{amssymb}
\usepackage{graphicx}

\numberwithin{equation}{section}
\definecolor{darkblue}{rgb}{0.02,0.548,0.908}
\definecolor{darkgreen}{rgb}{0,0.35,0}

\begin{document}

\title{Dynamical structure of Pure Lovelock gravity}
\small
\author{Naresh Dadhich$^{1,2}$\thanks{nkd@iucaa.ernet.in}, Remigiusz Durka$^{3}$\thanks{remigiusz.durka@ucv.cl},
Nelson Merino$^{3}$\thanks{nelson.merino@ucv.cl}, Olivera Miskovic$^{3}$\thanks{olivera.miskovic@pucv.cl}
\bigskip\\
\textit{{\small $^1$Centre for Theoretical Physics, Jamia Millia Islamia, }}\\
\textit{{\small New Delhi 110 025, India, and }}\\
\textit{{\small $^2$Inter-University Centre for Astronomy \& Astrophysics, }}\\\textit{{\small Post Bag 4 Pune 411 007, India}}\\
\textit{{\small $^3$Instituto de F\'{\i}sica, Pontificia Universidad Cat\'{o}lica de Valpara\'{\i}so, }}\\
\textit{{\small Casilla 4059, Valpara\'{\i}so, Chile}}}
\maketitle

\begin{abstract}
We study dynamical structure of Pure Lovelock gravity in spacetime dimensions
higher than four using the Hamiltonian formalism. The action consists of
cosmological constant and a single higher-order polynomial in the Riemann
tensor. Similarly to Einstein-Hilbert action, it possesses a unique constant
curvature vacuum and charged black hole solutions. We analyze physical degrees
of freedom and local symmetries in this theory. In contrast to the
Einstein-Hilbert case, a number of degrees of freedom depends on the
background and can vary from zero to the maximal value carried by the Lovelock theory.

\end{abstract}

\section{Introduction}

Lovelock-Lanczos gravity \cite{Lovelock:1971yv,Lanczos} is a natural
generalization of General Relativity to higher dimensions. It provides the
most general gravity action yielding the second order field equations in the
metric $g_{\mu\nu}(x)$. In a $(d+1)$-dimensional spacetime, the action is
given by%
\begin{equation}
I[g]=\int d^{d+1}x\sum\limits_{k=0}^{\left[  d/2\right]  }\alpha
_{k}\mathcal{L}_{k}\,.
\end{equation}
Each term in the sum is characterized by the coupling constant $\alpha_{k}$
multiplied by the dimensionally continued Euler density $\mathcal{L}_{k}$ of
order $k$ in the curvature,%
\begin{equation}
\mathcal{L}_{k}=\frac{1}{2^{k}}\,\sqrt{-g}\ \delta_{\nu_{1}...\nu_{2k}}%
^{\mu_{1}...\mu_{2k}}\ R_{\mu_{1}\mu_{2}}^{\nu_{1}\nu_{2}}\cdots R_{\mu
_{2k-1}\mu_{2k}}^{\nu_{2k-1}\nu_{2k}}\,.
\end{equation}
Here $R_{\ \beta\mu\nu}^{\alpha}$ is the Riemann curvature tensor and
$\delta_{\nu_{1}...\nu_{2k}}^{\mu_{1}...\mu_{2k}}$ is the totally
antisymmetric generalized Kronecker delta of order $k$ defined as the
determinant of the $k\times k$ matrix $\left[  \delta_{\nu_{1}}^{\mu_{1}%
}\delta_{\nu_{2}}^{\mu_{2}}\cdots\delta_{\nu_{k}}^{\mu_{k}}\right]  $. This
kind of action, polynomial in curvature, is of significant interest in
theoretical physics because it describes a wide class of models. It has been
shown in Refs.~\cite{Wheeler:1985nh,Wheeler:1985qd} that, for arbitrary
constants $\alpha_{k}$, a degeneracy may appear in the space of solutions
because the metric is not fully fixed by the field equations. For instance, if
the action has non-unique degenerate vacua, then the temporal component
$g_{tt}$ of any static spherically symmetric ansatz remains arbitrary
\cite{Dadhich:2012ma}. This problem can be avoided by a special choice of the
coefficients $\alpha_{k}$. The most simple example is given by the
Einstein-Hilbert (EH) term alone, which has the unique Minkowski vacuum.
Presence of the positive or negative cosmological constant term makes the
theory to have the unique de Sitter (dS) or anti-de Sitter (AdS) vacuum, respectively.

Another way to fix the coefficients $\alpha_{k}$ is to have a unique vacuum in
the theory but degenerated, which leads to Chern-Simons gravity in odd
dimensions and Born-Infeld gravity in even dimensions \cite{Troncoso:1999pk}.
In those theories all couplings are expressed only in terms of the
gravitational interaction and the cosmological constant. Also, choosing the
coefficients up to a certain order $k=1,\ldots,[d/2]\equiv N$ leads to a
family of non-equivalent theories whose black hole solutions were studied in
\cite{Crisostomo:2000bb} and also in \cite{Banados:1993ur} for the maximal
case with $k=N$.

Recently, there has been suggested another possibility, where instead of the
full Lovelock series only two terms in the sum are considered in the action:
the cosmological constant and a polynomial in the curvature of order $p$.
These Pure Lovelock (PL) gravities \cite{Cai:2006pq} remarkably admit
non-degenerate vacua in even dimensions, while in odd dimensions they have a
unique non degenerate dS and AdS vacuum. Their black hole solutions are
asymptotically indistinguishable from the ones appearing in General
Relativity \cite{Dadhich:2012ma}. That is even though the action and
equations of motion are free of the linear Einstein-Hilbert term. This similar asymptotic behavior of
two theories seems to extend also to the level of the dynamics and a number of physical degrees of freedom in the bulk.

The properties of PL gravity have been discussed in the literature.
 Stability of PL black holes has been analyzed in
\cite{Gannouji:2013eka}. Application of gauge/gravity duality to phase
transitions in quantum field theories dual to Pure Gauss-Bonnet AdS gravity
were studied in Ref.~\cite{Aranguiz:2015voa}. It can be shown that in any
dimension $d+1$ there is a special power $p$ such that the black hole entropy
behaves as in any particular lower dimension. In case of the maximum power,
$p=N$, such as five-dimensional Pure Gauss-Bonnet action, they exhibit a
peculiar thermodynamical behavior \cite{Dadhich:2012ma,Dadhich:2012eg}, where
temperature and entropy bear the same relation to horizon radius as
in the case for $3D$ and $4D$ dimensions, respectively. Thermodynamical parameters are
thus universal in terms of horizon radius for all odd $D=2N+1$ and even $D=2N+2$
dimensions.

Dynamical aspects of PL theory were analyzed in Ref.~\cite{Camanho:2015hea} in
terms of analogs of the Riemann and Weyl tensors for $N$th order PL
gravity. It turns out that it is possible to define an $N$th order Riemann
curvature with the property that trace of its Bianchi derivative yields the same
divergence free (analogue of Einstein tensor) second rank tensor as the one obtained by
the corresponding Lovelock polynomial action. Thus, one can obtain the gravitational equations
for PL gravity \cite{Dadhich:2012eqn, Dadhich:2015dp} in the same way as one does for
the Einstein equations from the Bianchi identity. However, there is one crucial difference,
which is that the second Bianchi identity (i.e., vanishing of Bianchi derivative) is only satisfied by the Riemann
tensor and not by its $N$th order analogue. The former has therefore a direct link to the metric,
while for the latter this relation is more involved. What yields the divergence-free tensor
is vanishing of the trace of Bianchi derivative, and not necessarily
derivative itself. From this perspective, PL gravity could be seen as kinematic, which means that the $N$th order Riemann tensor is entirely given in terms of the corresponding
Ricci tensor in all critical odd $D=2N+1$ dimensions, and it becomes dynamic in the even $D=2N+2$
dimensions. This might uncover a universal feature of gravitational dynamics in all critical odd and even dimensions,
making it drastically different in critical odd dimensions.
More precisely, the PL vacuum is flat with respect to $N$th order Riemann tensor, but not relative to Riemann tensor. This suggests that there
are no dynamical degrees of freedom in the critical odd dimensions relative to the former but that may not be the case for the latter.

On the other hand it has been argued in Ref.~\cite{Teitelboim} that the metric Lovelock theory
should have the same number of degrees of freedom as the higher-dimensional
Einstein-Hilbert gravity, namely $D(D-3)/2$. This is different than expected from our previous discussion, which suggested fewer physical fields.
However, a number of degrees of freedom can change with the backgrounds. For example, Lovelock-Chern-Simons gravity
has different number of degrees of freedom in different sectors of the phase
space \cite{Banados-Garay-Henneaux5,Banados-Garay-HenneauxD}. Due to
non-linearity of the theory, the symplectic matrix might have different rank
depending on the background \cite{Saavedra:2000wk} causing more symmetries and
less degrees of freedom in some of them, what was explicitly demonstrated in
Chern-Simons supergravity \cite{Giribet-Merino-Miskovic}. It can also happen that the constraints become functionally depended in
certain symmetric backgrounds \cite{Miskovic:2003ex}.

We wish therefore to
provide a detailed analysis of the dynamical structure of PL theory by
explicitly performing Hamiltonian analysis and exploring until what extent it is
similar to General Relativity, and whether it exhibits any additional universal features.

\section{Pure Lovelock gravity}

We focus on Pure Lovelock gravity of order $p$ in $(d+1)$-dimensions, whose
action consists of the unique Lovelock term, $\mathcal{L}_{p}$, and
the cosmological constant $\mathcal{L}_{0}$,
\begin{equation}
I[g]=-\kappa\int d^{d+1}x\sqrt{-g}\left(  \frac{1}{2^{p}}\ \delta_{\nu
_{1}...\nu_{2p}}^{\mu_{1}...\mu_{2p}}\,R_{\mu_{1}\mu_{2}}^{\nu_{1}\nu_{2}%
}\cdots R_{\mu_{2p-1}\mu_{2p}}^{\nu_{2p-1}\nu_{2p}}-2\Lambda\right)  \,,
\label{PL}%
\end{equation}
where $\alpha_{p}=-\kappa$ and $\alpha_{0}=2\kappa\Lambda$. The gravitational
constant $\kappa$ has dimension (length)$^{d+1-2p}$ and the cosmological
constant has dimension of (length)$^{-2p}$, and not (length)$^{-2}$ as in
General Relativity. Varying the action with respect to the metric $g_{\mu\nu
}(x)$, one obtains equations of motion in the form%
\begin{equation}
^{(p)}G_{\nu}^{\mu}+\Lambda\delta_{\nu}^{\mu}=0\ , \label{PL_metric_fe}%
\end{equation}
where $\Lambda=0$ or $\Lambda=\dfrac{\left(  \pm1\right)  ^{p}d!}{2\left(
d-2p\right)  !\ell^{2p}}$, and generalized Einstein tensor is symmetric of $p
$-th order in the curvature,%
\begin{equation}
^{(p)}G_{\nu}^{\mu}=-\frac{1}{2^{p+1}}\,\delta_{\nu\mu_{1}...\mu_{2p}}^{\mu
\nu_{1}...\nu_{2p}}\,R_{\nu_{1}\nu_{2}}^{\mu_{1}\mu_{2}}\cdots R_{\nu
_{2p-1}\nu_{2p}}^{\mu_{2p-1}\mu_{2p}}\,.
\end{equation}
The form of $\Lambda$ given above is such that $\Lambda=0$ has Minkowski
metric as a particular solution, whereas $\Lambda\neq0$ has the dS (sign $+ $)
and AdS (sign $-$) space of the radius $\ell$ as solutions of the PL field equations.

Due to the presence of local symmetries in the theory, not all components of
the metric are physical. In order to determine dynamically propagating fields
in the bulk, we turn to the Hamiltonian formalism, which provides a systematic
method to separate physical variables from the non-physical ones. However,
applying the canonical analysis to PL action in the metric formalism is
technically involved, even though it depends on velocities only. A reason is
that it is higher-order in the curvature.

On the other hand, if we write the action (\ref{PL}) in Palatini formalism
$\tilde{I}[g,\Gamma]$, where the metric $g_{\mu\nu}$ and affine connection
$\Gamma_{\mu\nu}^{\lambda}$ are treated as independent fundamental fields,
then the theory naturally includes torsional degrees of freedom. Then the
vanishing torsion would correspond to just a particular solution of the fields
equations, whereas in General Relativity it is the only solution. A wider
space of solutions can be avoided by introducing a Lagrange multiplier that
forces the torsion to vanish, in a way that field equations become the ones of
PL gravity in Riemann space.

In the next section, we reformulate the PL gravity in first order formalism,
linear in velocities, which makes it much simpler to apply the Hamiltonian analysis.

\subsection{First order formalism}

The fundamental fields in the first order formalism, vielbein $e_{\mu}^{a}\left(
x\right)  $ and spin-connection $\omega_{\mu}^{ab}\left(  x\right)  $ are
related to the fields in the tensorial formalism through the relations
$g_{\mu\nu}=\eta_{ab}\,e_{\mu}^{a}e_{\nu}^{b}$ and $\Gamma_{\mu\nu}^{\lambda
}=\omega_{\nu}^{ab}e_{a}^{\lambda}e_{a\mu}+e_{a}^{\lambda}\partial_{\nu}%
e_{\mu}^{a}$, where $a,b=0,1,\ldots,d$ are the Lorentz indices. Note that the
change of variables $(g,\Gamma)\rightarrow(e,\omega)$ is not unique, but is
determined up to Lorentz rotations. With the new fields, we obtain Riemann
curvature tensor $R_{\mu\nu}^{ab}$ and torsion tensor $T_{\mu\nu}^{a}$ as
\begin{align}
R_{\mu\nu}^{ab}  &  =\partial_{\mu}\omega_{\nu}^{ab}-\partial_{\nu}\omega
_{\mu}^{ab}+\omega_{\mu b}^{a}\omega_{\nu}^{bc}-\omega_{\nu b}^{a}\omega_{\mu
}^{bc}\,,\nonumber\\
T_{\mu\nu}^{a}  &  =D_{\mu}e_{\nu}^{a}-D_{\nu}e_{\mu}^{a}\,,
\end{align}
where $D=D(\omega)$ is a covariant derivative with respect to the spin
connection acting on the Lorentz indices only, e.g., $D_{\mu}e_{\nu}%
^{a}=\partial_{\mu}e_{\nu}^{a}+\omega_{\mu b}^{a}e_{\nu}^{b}$ .

Naively, the first order PL action can be cast in the form%
\begin{equation}
\tilde{I}\left[  e,\omega\right]  =\int d^{d+1}x\ \left(  \alpha
_{0}\mathcal{L}_{0}+\alpha_{p}\mathcal{L}_{p}\right)  \,, \label{PL_for_p}%
\end{equation}
where we rescaled $\alpha_{k}\mathcal{\rightarrow-}\frac{\alpha_{k}%
}{(d+1-2k)!}$ and $\mathcal{L}_{k}\rightarrow-(d+1-2k)!\mathcal{L}_{k}$, and
the Euler densities now become polynomials in $R$ and $e$,
\begin{align}
\mathcal{L}_{0}  &  =\epsilon_{a_{1}\cdots a_{d+1}}\epsilon^{\mu_{1}\cdots
\mu_{d+1}}\,e_{\mu_{1}}^{a_{1}}\cdots e_{\mu_{d+1}}^{a_{d+1}}\sim
e^{d+1}\,,\nonumber\\
\mathcal{L}_{p}  &  =\frac{1}{2^{p}}\,\epsilon_{a_{1}\cdots a_{d+1}}%
\epsilon^{\mu_{1}\cdots\mu_{d+1}}\,R_{\mu_{1}\mu_{2}}^{a_{1}a_{2}}\cdots
R_{\mu_{2p-1}\mu_{2p}}^{a_{2p-1}a_{2p}}e_{\mu_{2p+1}}^{a_{2p+1}}\cdots
e_{\mu_{d+1}}^{a_{d+1}}\sim R^{p}e^{d+1-2p}\,.
\end{align}
Notation for the Levi-Civita symbol $\epsilon^{\mu_{1}\cdots\mu_{d+1}}$ is
given in Appendix \ref{Conv}.\ The coupling constants become%
\begin{equation}
\alpha_{0}=\frac{2\Lambda\kappa}{(d+1)!}\,,\qquad\alpha_{p}=-\frac{\kappa
}{(d+1-2p)!}\,.
\end{equation}
However, the field equations obtained from the action (\ref{PL_for_p}) after
varying it in $e_{\mu}^{a}$ and $\omega_{\mu}^{ab}$ are, respectively,
\begin{align}
0  &  =\epsilon_{aa_{1}\cdots a_{d}}\epsilon^{\mu\mu_{1}\cdots\mu_{d}}\left(
\frac{1}{2^{p}}\,R_{\mu_{1}\mu_{2}}^{a_{1}a_{2}}\cdots R_{\mu_{2p-1}\mu_{2p}%
}^{a_{2p-1}a_{2p}}e_{\mu_{2p+1}}^{a_{2p+1}}\cdots e_{\mu_{d}}^{a_{d}}%
+\frac{\alpha_{0}\,(d+1)}{\alpha_{p}(d+1-2p)}\,e_{\mu_{1}}^{a_{1}}\cdots
e_{\mu_{d}}^{a_{d}}\right)  ,\label{fe1_PL}\\
0  &  =\epsilon_{aba_{2}\cdots a_{d}}\epsilon^{\mu\mu_{1}\cdots\mu_{d}%
}\,\left(  \frac{1}{2^{p}}\,R_{\mu_{2}\mu_{3}}^{a_{2}a_{3}}\cdots
R_{\mu_{2p-2}\mu_{2p-1}}^{a_{2p-2}a_{2p-1}}\ T_{\mu_{2p}\mu_{1}}^{a_{2p}%
}\ e_{\mu_{2p+1}}^{a_{2p+1}}\cdots e_{\mu_{d}}^{a_{d}}\right)  \,.
\label{fe2_PL}%
\end{align}
These equations are not equivalent to the PL field Eqs.~(\ref{PL_metric_fe})
because the Riemann spaces for which $T_{\mu\nu}^{a}=0$ are not the only
solutions of Eqs.~(\ref{fe2_PL}) when $d+1>4$ and $p>1$. Thus, treating
$\left(  e_{\mu}^{a},\omega_{\mu}^{ab}\right)  $ as independent fields changes
the dynamics of the system. In order to use first order formalism and, at the
same time, obtain field equations of Pure Lovelock gravity where $T_{\mu\nu
}^{a}=0$ is the unique solution, we introduce a Lagrange multiplier
$\lambda_{a}^{\mu\nu}$ that forces the torsion tensor to vanish through a
constraint. The new action reads%
\begin{equation}
I[e,\omega,\lambda]=\int d^{d+1}x\ \left(  \alpha_{0}\mathcal{L}_{0}%
+\alpha_{p}\mathcal{L}_{p}+\frac{1}{2}\ T_{\mu\nu}^{a}\,\lambda_{a}^{\mu\nu
}\right)  \,. \label{PL_with_multiplier}%
\end{equation}
The field $\lambda_{a}^{\mu\nu}(x)$ is antisymmetric in the indices $\left[
\mu\nu\right]  $.

Although proposed action is explicitly torsionless, it does not imply that the
equations\ of motion give the dynamics equivalent to the PL one. An example of
the system where an addition of the constraint $T^{a}\lambda_{a}$ modifies the
dynamics of a theory is Topologically Massive Gravity, where it introduces a
term involving the Cotton tensor
\cite{Deser:1991qk,Carlip:2008qh,Blagojevic:2008bn}. There, the term with the
multiplier has nontrivial implications on derivation of conserved charges
\cite{Miskovic:2009kr}. An influence of the multiplier, therefore, has to be
well-understood on the level of the field equations.

The action (\ref{PL_with_multiplier}) reaches an extremum on the equations of
motion,%
\begin{align}
\delta e_{\mu}^{a}  &  :0=\epsilon_{aa_{1}\cdots a_{d}}\epsilon^{\mu\mu
_{1}\cdots\mu_{d}}\left[  \frac{\alpha_{p}}{2^{p}}\,(d+1-2p)\,R_{\mu_{1}%
\mu_{2}}^{a_{1}a_{2}}\cdots R_{\mu_{2p-1}\mu_{2p}}^{a_{2p-1}a_{2p}}%
\,e_{\mu_{2p+1}}^{a_{2p+1}}\cdots e_{\mu_{d}}^{a_{d}}\right. \nonumber\\
&  \qquad+\left.  \alpha_{0}\,(d+1)\,e_{\mu_{1}}^{a_{1}}\cdots e_{\mu_{d}%
}^{a_{d}}\rule{0pt}{15pt}\right]  +D_{\nu}\lambda_{a}^{\mu\nu}%
\ ,\label{field_eq1}\\
\delta\omega_{\mu}^{ab}  &  :0=\frac{1}{2^{p}}\,\epsilon_{aba_{2}\cdots a_{d}%
}\epsilon^{\mu\mu_{1}\cdots\mu_{d}}\,R_{\mu_{2}\mu_{3}}^{a_{2}a_{3}}\cdots
R_{\mu_{2p-2}\mu_{2p-1}}^{a_{2p-2}a_{2p-1}}\ T_{\mu_{2p}\mu_{1}}^{a_{2p}%
}\ e_{\mu_{2p+1}}^{a_{2p+1}}\cdots e_{\mu_{d}}^{a_{d}}\nonumber\\
&  \qquad+\frac{1}{2}\,\left(  e_{b\nu}\lambda_{a}^{\mu\nu}-e_{a\nu}%
\lambda_{b}^{\mu\nu}\right)  \,,\label{field_eq2}\\
\delta\lambda_{a}^{\mu\nu}  &  :0=T_{\mu\nu}^{a}\,. \label{field_eq3}%
\end{align}
In addition, the curvature and torsion tensors satisfy the First and Second
Bianchi identities,%
\begin{align}
D_{\mu}T_{\rho\sigma}^{a}+D_{\rho}T_{\sigma\mu}^{a}+D_{\sigma}T_{\mu\rho}^{a}
&  =R_{\ \ \mu\rho}^{ab}e_{b\sigma}+R_{\ \ \rho\sigma}^{ab}e_{b\mu
}+R_{\ \ \sigma\mu}^{ab}e_{b\rho}\ ,\nonumber\\
D_{\mu}R_{\ \ \rho\sigma}^{ab}+D_{\rho}R_{\ \ \sigma\mu}^{ab}+D_{\sigma
}R_{\ \ \mu\rho}^{ab}  &  =0\,. \label{Bianchi 1}%
\end{align}
When the torsion tensor vanishes, the field equation (\ref{field_eq2})
becomes
\begin{equation}
0=e_{b\nu}\lambda_{a}^{\mu\nu}-e_{a\nu}\lambda_{b}^{\mu\nu}\ ,
\label{multiplier_eq}%
\end{equation}
from where $d(d+1)^{2}/2$ components of $\lambda_{a}^{\mu\nu}$ can be solved
as%
\begin{equation}
\lambda_{a}^{\mu\nu}=0\,. \label{multiplier_equation}%
\end{equation}
This result is obtained by rewriting (\ref{multiplier_eq}) with the Lorentz
indices as $\lambda_{a,bc}-\lambda_{c,ba}=0$, and combining it with two other
expressions obtained by performing the permutation of indices, which directly
leads to $\lambda_{a,bc}=0$ and therefore (\ref{multiplier_equation}). Using
(\ref{multiplier_equation}), the last equation (\ref{field_eq1}) is indeed
equivalent to the Lovelock field equations in Riemann space. The Bianchi
identities (\ref{Bianchi 1}) in that case read%
\begin{equation}
R_{\ \ \left(  \sigma\mu\rho\right)  }^{a}=0\ ,\qquad D_{(\mu}R_{\rho\sigma
)}^{ab}=0\,. \label{Bianchi 2}%
\end{equation}

\section{Action in the time-like foliation}

Hamiltonian formalism is not explicitly covariant because it presents all the
quantities in the time-like foliation $x^{\mu}=\left(  t,x^{i}\right)  $,
where $x^{0}=t\in\mathbb{R}$ is the temporal coordinate and $x^{i}$
($i=1,\ldots,d$) are local coordinates at the spatial section $\Sigma$.

In the tangent space, we decompose the indices as $a=(0,\bar{a})$. The
vielbein $e_{\mu}^{a}$ is invertible on $\mathbb{R}\times\Sigma$ and its
inverse is $e_{a}^{\mu}$. We require that $e_{0}^{t}\neq0$ and that the
$d$-dimensional vielbein $e_{i}^{\bar{a}}$ is also invertible with the inverse%
\begin{equation}
^{(d)}e_{\bar{a}}^{i}=e_{\bar{a}}^{i}-\frac{e_{0}^{i}e_{\bar{a}}^{t}}%
{e_{0}^{t}}\ . \label{inverse (d)e}%
\end{equation}

In order to introduce canonical variables in the action
(\ref{PL_with_multiplier}), we have to define the action in configurational
space, that is, in terms of the fields $e_{\mu}^{a}$, $\omega_{\mu}^{ab}$ and
its velocities $\dot{e}_{\mu}^{a}$, $\dot{\omega}_{\mu}^{ab}$. To this end, we
have the splitting of the fields in the time-like foliation
\[
e_{\mu}^{a}\rightarrow\left(  e_{t}^{a},e_{i}^{a}\right)  \,,\qquad\omega
_{\mu}^{ab}\rightarrow\left(  \omega_{t}^{ab},\omega_{i}^{ab}\right)  \,,
\]
and similarly for the multiplier $\lambda_{a}^{\mu\nu}\rightarrow(\lambda
_{a}^{ti}\equiv\lambda_{a}^{i},\lambda_{a}^{ij})$. It is worthwhile noticing
that $\omega_{t}^{ab}$ transforms as a tensor of rank 2 under local Lorentz
transformations on $\Sigma$ and $\omega_{i}^{ab}$ as the Lorentz gauge connection.

Since $L=\int d^{d}x\,\mathcal{L}$, the Lagrangian scalar density of
(\ref{PL_with_multiplier}) can be written in a compact way,%
\begin{equation}
\mathcal{L}=\frac{1}{2}\,\dot{\omega}_{i}^{ab}\,\mathcal{L}_{ab}^{i}+\dot
{e}_{i}^{a}\,\lambda_{a}^{i}+\frac{1}{2}\,\omega_{t}^{ab}\,\mathcal{S}%
_{ab}+e_{t}^{a}\,\mathcal{S}_{a}+\frac{1}{2}\,T_{ij}^{a}\,\lambda_{a}^{ij}\,.
\label{L_density}%
\end{equation}
We neglect all boundary terms. In the action above, we introduce the
quantities which do not depend on velocities and time-like components,%
\begin{align}
\mathcal{L}_{ab}^{i}  &  =\frac{p\alpha_{p}}{2^{p-2}}\,\epsilon_{aba_{2}\cdots
a_{d}}\epsilon^{ii_{2}\cdots i_{d}}R_{i_{2}i_{3}}^{a_{2}a_{3}}\cdots
R_{i_{2p-2}i_{2p-1}}^{a_{2p-2}a_{2p-1}}e_{i_{2p}}^{a_{2p}}\cdots e_{i_{d}%
}^{a_{d}}\,,\label{Lab_density}\\
\mathcal{S}_{a}  &  =\mathcal{H}_{a}+D_{i}\lambda_{a}^{i}\,,
\label{Sa_density}\\
\mathcal{S}_{ab}  &  =\mathcal{H}_{ab}+e_{bi}\lambda_{a}^{i}-e_{ai}\lambda
_{b}^{i}\,, \label{Sab_density}%
\end{align}
where%
\begin{align}
\mathcal{H}_{a}  &  =\epsilon_{aa_{1}\cdots a_{d}}\epsilon^{i_{1}\cdots i_{d}%
}\left[  (d+1)\,\alpha_{0}\,e_{i_{1}}^{a_{1}}\cdots e_{i_{d}}^{a_{d}}%
+\frac{\alpha_{p}}{2^{p}}\,(d+1-2p)\,R_{i_{1}i_{2}}^{a_{1}a_{2}}\cdots
R_{i_{2p-1}i_{2p}}^{a_{2p-1}a_{2p}}e_{i_{2p+1}}^{a_{2p+1}}\cdots e_{i_{d}%
}^{a_{d}}\right]  \,,\nonumber\\
\mathcal{H}_{ab}  &  =\frac{p\alpha_{p}}{2^{p-1}}\,(d+1-2p)\,\epsilon
_{aba_{2}\cdots a_{d}}\,\epsilon^{i_{1}\cdots i_{d}}\,R_{i_{2}i_{3}}%
^{a_{2}a_{3}}\cdots R_{i_{2p-2}i_{2p-1}}^{a_{2p-2}a_{2p-1}}T_{i_{1}i_{2p}%
}^{a_{2p}}e_{i_{2p+1}}^{a_{2p+1}}\cdots e_{i_{d}}^{a_{d}}\,.
\end{align}
The Lagrangian (\ref{L_density}) is similar to the one of Chern-Simons theory,
whose Hamiltonian analysis was studied in Ref.~\cite{Banados-Garay-HenneauxD}.

\section{Hamiltonian analysis in five dimensions}

Let us start with the simplest case of five-dimensional Pure Gauss-Bonnet
action ($d=4$, $p=2$),%
\begin{equation}
I=\int d^{5}x\left[  \epsilon_{abcde}\epsilon^{\mu\nu\rho\sigma\gamma}\left(
\alpha_{0}\,e_{\mu}^{a}e_{\nu}^{b}e_{\rho}^{c}e_{\sigma}^{d}e_{\gamma}%
^{e}+\frac{\alpha_{2}}{4}\,R_{\mu\nu}^{ab}R_{\rho\sigma}^{cd}e_{\gamma}%
^{e}\right)  +\frac{1}{2}\,T_{\mu\nu}^{a}\lambda_{a}^{\mu\nu}\right]  \,.
\end{equation}
The Lagrangian has the form (\ref{L_density}) with particular tensors%
\begin{align}
\mathcal{L}_{ab}^{i}  &  =2\alpha_{2}\,\epsilon^{ijkl}\epsilon_{abcde}%
\,R_{jk}^{cd}e_{l}^{e}\,,\nonumber\\
\mathcal{S}_{ab}  &  =\mathcal{H}_{ab}+e_{bi}\lambda_{a}^{i}-e_{ai}\lambda
_{b}^{i}\,,\nonumber\\
\mathcal{S}_{a}  &  =\mathcal{H}_{a}+D_{i}\lambda_{a}^{i}\,,\nonumber\\
\mathcal{H}_{ab}  &  =\alpha_{2}\,\epsilon_{abcde}\,\epsilon^{ijkl}%
\,R_{ij}^{cd}T_{kl}^{e}\,,\nonumber\\
\mathcal{H}_{a}  &  =\epsilon_{abcde}\epsilon^{ijkl}\left(  5\alpha_{0}%
\,e_{i}^{b}e_{j}^{c}e_{k}^{d}e_{l}^{e}+\frac{\alpha_{2}}{4}\,R_{ij}^{bc}%
R_{kl}^{de}\right)  \,,
\end{align}
and the multipliers are conveniently written as%
\begin{equation}
\lambda_{a}^{i}=\frac{1}{3!}\,\epsilon^{ijkl}\lambda_{a,jkl}\,,\qquad
\lambda_{a}^{ij}=\frac{1}{2!}\,\epsilon^{ijkl}\lambda_{a,tkl}\,.
\end{equation}
If we denote the generalized coordinates by $q^{M}(x)$ and the corresponding
conjugated momenta by $\pi_{M}(x)$,
\begin{equation}
q^{M}=\{e_{t}^{a},e_{i}^{a},\omega_{t}^{ab},\omega_{i}^{ab},\lambda_{a}%
^{i},\lambda_{a}^{ij}\}\,,\qquad\pi_{M}=\{\pi_{a}^{t},\pi_{a}^{i},\pi_{ab}%
^{t},\pi_{ab}^{i},p_{i}^{a},p_{ij}^{a}\}\,, \label{generalized_coord}%
\end{equation}
we can use the definition $\pi_{M}=\frac{\partial\mathcal{L}}{\partial\dot
{q}^{M}}$ to find $\pi_{ab}^{i}=\mathcal{L}_{ab}^{i}$ and $\pi_{a}^{i}%
=\lambda_{a}^{i}$, while all other momenta are zero. Thus, the Hessian matrix
$\frac{\partial^{2}\mathcal{L}}{\partial\dot{q}^{M}\partial\dot{q}^{N}}$ is
not invertible and we cannot express all velocities in terms of the momenta.
In turn, we get the constraints, called%
\begin{equation}
\text{Primary constraints\textit{:}}\qquad\Phi_{M}=\{\phi_{a}^{t},\phi_{a}%
^{i},\phi_{ab}^{t},\phi_{ab}^{i},p_{i}^{a},p_{ij}^{a}\}\,.
\label{Primary_const}%
\end{equation}
They are defined on the phase space as
\begin{equation}%
\begin{array}
[c]{llll}%
\phi_{a}^{t} & =\pi_{a}^{t}\approx0\,,\qquad & \phi_{a}^{i} & =\pi_{a}%
^{i}-\lambda_{a}^{i}\approx0\,,\\
\phi_{ab}^{t} & =\pi_{ab}^{t}\approx0\,, & \phi_{ab}^{i} & =\pi_{ab}%
^{i}-\mathcal{L}_{ab}^{i}\approx0\,,\\
p_{i}^{a} & \approx0\,, & p_{ij}^{a} & \approx0\,.
\end{array}
\label{Primary_const2}%
\end{equation}
The surface $\Phi_{M}\approx0$ in the phase space is called the primary
constraint surface, $\Gamma_{P}$. The weak equality $f(q,\pi)\approx0$ on
$\Gamma_{P}$ implies that a phase space function $f$ vanish on $\Gamma_{P}$,
but its derivatives (variations) are non-vanishing. This is different than the
strong equality, $f(q,\pi)=0$, where both $f$ and its variations vanish on
$\Gamma_{P}$. This distinction is relevant for definition of Poisson brackets,
since $f\approx0$ does not imply $\{f,\cdots\}\approx0$.

To simplify notation, we write the arguments of the phase space functions
symbolically, assuming that all quantities are defined at the same instant,
$x^{0}=x^{\prime0}=t$,%
\begin{align}
A  &  =A\left(  x\right)  \,,\qquad B^{\prime}=B\left(  x^{\prime}\right)
\,,\qquad\partial_{i}=\frac{\partial}{\partial x^{i}}\,,\qquad\partial
_{i}^{\prime}=\frac{\partial}{\partial x^{\prime i}}\,,\nonumber\\
\delta &  =\delta\left(  \vec{x}-\vec{x}^{\prime}\right)  \,,\qquad\delta
_{cd}^{ab}=\delta_{c}^{a}\delta_{d}^{b}-\delta_{d}^{a}\delta_{c}^{b}\,.
\label{notation}%
\end{align}

The fundamental Poisson brackets (PBs) different than zero are
\begin{align}
\{e_{\mu}^{a},\pi_{b}^{\prime\nu}\}  &  =\delta_{b}^{a}\,\delta_{\mu}^{\nu
}\,\delta\,,\nonumber\\
\{\omega_{\mu}^{ab},\pi_{cd}^{\prime\nu}\}  &  =\delta_{cd}^{ab}\,\delta_{\mu
}^{\nu}\,\delta\,,\nonumber\\
\{\lambda_{a}^{i},p_{j}^{\prime b}\}  &  =\delta_{b}^{a}\,\delta_{j}%
^{i}\,\delta\,,\nonumber\\
\{\lambda_{a}^{ij},p_{kl}^{\prime b}\}  &  =\delta_{b}^{a}\,\delta_{kl}%
^{ij}\,\delta\,.
\end{align}
The symplectic matrix $\Omega_{MN}$ of the primary constraints reads%
\begin{equation}
\{\Phi_{M},\Phi_{N}^{\prime}\}=\Omega_{MN}\,\delta\,,
\end{equation}
and it is antisymmetric, $\Omega_{MN}=-\Omega_{NM}$. The only (independent)
submatrices of the symplectic matrix different than zero are%
\begin{align}
\{\phi_{ab}^{i},\phi_{cd}^{\prime j}\}  &  =\Omega_{abcd}^{ij}\,\delta
=-8\alpha_{2}\,\epsilon^{ijkl}\epsilon_{abcde}\,T_{kl}^{e}\,\delta
\ ,\nonumber\\
\{\phi_{ab}^{i},\phi_{c}^{\prime j}\}  &  =\Omega_{abc}^{ij}\,\delta
=-2\alpha_{2}\epsilon^{ijkl}\epsilon_{abcde}\,R_{kl}^{de}\,\delta
\,,\nonumber\\
\{\phi_{a}^{i},p_{j}^{\prime b}\}  &  =-\delta_{a}^{b}\delta_{j}^{i}%
\,\delta\,. \label{symplectic}%
\end{align}
The canonical Hamiltonian, $\mathcal{H}_{C}=\pi_{M}\dot{q}^{M}-\mathcal{L}$,
defined on $\Gamma_{P}$ is%
\begin{equation}
\mathcal{H}_{C}(p,q)=-\frac{1}{2}\,\omega_{t}^{ab}\,\mathcal{S}_{ab}-e_{t}%
^{a}\,\mathcal{S}_{a}-\frac{1}{2}\,T_{ij}^{a}\,\lambda_{a}^{ij}\,,
\label{Canonical_5D}%
\end{equation}
and the total Hamiltonian, defined on the full phase space $\Gamma$, is
obtained by introducing the indefinite multipliers $u^{M}(x)$,
\begin{equation}
\mathcal{H}_{T}(p,q,u)=\mathcal{H}_{C}(p,q)+u^{M}\Phi_{M}(p,q)\,,
\label{Total_5D}%
\end{equation}
where $u^{M}=\{u_{t}^{a},u_{i}^{a},u_{t}^{ab},u_{i}^{ab},v_{a}^{i},v_{a}%
^{ij}\}$. Evolution of any quantity $A(q(x),\pi(x))=A(x)$ in the phase space
is given by%
\begin{align}
\dot{A}  &  =\int d\vec{x}^{\prime}\left(  \left\{  A,\mathcal{H}_{C}^{\prime
}\right\}  +u^{\prime M}\left\{  A,\Phi_{M}^{\prime}\right\}  \right)
\nonumber\\
&  \approx\int d\vec{x}^{\prime}\left\{  A,\mathcal{H}_{T}^{\prime}\right\}
\,. \label{evolution}%
\end{align}
This allows us to identify some field velocities with the Hamiltonian
multipliers,
\begin{equation}%
\begin{array}
[c]{ll}%
\dot{\omega}_{i}^{ab}=u_{i}^{ab}\,,\qquad & \dot{e}_{i}^{a}=u_{i}^{a}\,,\\
\dot{\lambda}_{a}^{ij}=v_{a}^{ij}\,, & \dot{\lambda}_{a}^{i}\,=v_{a}^{i}\,.
\end{array}
\end{equation}

Consistency of the theory requires that the primary constraints remain on the
constraint surface during their evolution, that is,%
\begin{equation}
\dot{\Phi}_{M}=\int d\vec{x}^{\prime}\{\Phi_{M},\mathcal{H}_{C}^{\prime
}\}+\Omega_{MN}\,u^{N}\approx0\,. \label{evol_eq2}%
\end{equation}
These consistency conditions will either solve some multipliers, or lead to
the secondary constraints, or will be identically satisfied.

When the symplectic matrix has zero modes and $\{\Phi_{M},\mathcal{H}%
_{C}\}\neq0$, the consistency conditions lead to the secondary constraints,%
\begin{align}
\dot{\phi}_{a}^{t}  &  =\mathcal{S}_{a}\approx0\,,\label{sec_const_1}\\
\dot{\phi}_{ab}^{t}  &  =\mathcal{S}_{ab}\approx0\,,\label{sec_const_2}\\
\dot{p}_{ij}^{a}  &  =T_{ij}^{a}\approx0\,. \label{sec_const_3}%
\end{align}
Other consistency conditions solve Hamiltonian multipliers, such as $\dot
{p}_{i}^{a}\approx0$, which gives
\begin{equation}
u_{i}^{a}=D_{i}e_{t}^{a}-\omega_{t}^{ab}e_{bi}\ . \label{u_i_a}%
\end{equation}
On the other hand, from $\dot{\phi}_{a}^{i}\approx0$ we solve the multiplier,%
\begin{equation}
v_{a}^{i}=-\epsilon_{abcde}\epsilon^{ijkl}\left[  20\alpha_{0}\,e_{t}^{b}%
e_{j}^{c}e_{k}^{d}e_{l}^{e}+\alpha_{2}\,R_{kl}^{de}\left(  u_{j}^{bc}%
-D_{j}\omega_{t}^{bc}\right)  \right]  +\omega_{t\,a}^{b}\lambda_{b}^{i}%
+D_{j}\lambda_{a}^{ij}\ . \label{V_i_a}%
\end{equation}
Using the Bianchi identities, $D_{j}\epsilon_{abcde}=0$ and the property\ that
any totally antisymmetric tensor of rank 6 defined in five dimensions must
vanish, that is,
\[
-\epsilon_{bcdef}\,e_{aj}+\epsilon_{cdefa}e_{bj}-\epsilon_{defab}%
\,e_{cj}+\epsilon_{efabc}\,e_{dj}-\epsilon_{fabcd}\,e_{ej}+\epsilon
_{abcde}\,e_{fj}=0\,,
\]
the last consistency condition for $\phi_{ab}^{i}$ becomes
\begin{equation}
0\approx\dot{\phi}_{ab}^{i}\approx e_{at}\lambda_{b}^{i}-e_{bt}\lambda_{a}%
^{i}-\lambda_{a}^{ij}e_{bj}+\lambda_{b}^{ij}e_{aj}\,. \label{consistency(ab)}%
\end{equation}
One can show, in a similar way as for Eq.~(\ref{multiplier_equation}), that
the constraints (\ref{sec_const_2}) and\ (\ref{consistency(ab)}) are now
equivalent to zero multipliers $\lambda_{a}^{i}\approx0$ and $\lambda_{a}%
^{ij}\approx0$.

So far, we have found the following
\begin{equation}
\text{Secondary constraints:}\qquad\mathcal{S}_{a}\approx0\,,
\quad T_{ij}^{a}\approx0\,,\quad\lambda_{a}^{i}\approx0\,,\quad\lambda_{a}^{ij}\approx0\,,\label{Secondary}
\end{equation}
and we determined the multipliers $u_{i}^{a}$ and $v_{a}^{i}$. The functions
$\{u_{t}^{a},u_{t}^{ab},u_{i}^{ab},v_{a}^{ij}\}$ remain arbitrary. A
submanifold $\Gamma_{S}\subset\Gamma$ defines the secondary constraint
surface, where all constraints discovered until now vanish.

To ensure that the secondary constraints $\lambda$ evolve on the constraint
surface $\Gamma_{S}$, we require that $\dot{\lambda}_{a}^{i}=v_{a}^{i}$ and
$\dot{\lambda}_{a}^{ij}=v_{a}^{ij}$ vanish. It leads to $v_{a}^{i}=0$, which
by Eq.~(\ref{V_i_a}) can be\textbf{\ }equivalently expressed as%
\begin{equation}
\chi_{a}^{i}=-\epsilon_{abcde}\epsilon^{ijkl}\left[  20\alpha_{0}\,e_{t}%
^{b}e_{j}^{c}e_{k}^{d}e_{l}^{e}+\alpha_{2}\,R_{kl}^{de}\left(  u_{j}%
^{bc}-D_{j}\omega_{t}^{bc}\right)  \right]  \approx0\quad\text{and}\quad
v_{a}^{ij}=0\ . \label{v_inconclusive}%
\end{equation}

Before we continue, we can notice that the pairs of conjugated variables
$(\lambda,p)$, all being the constraints, have their PB's whose r.h.s
(symplectic form) is invertible on $\Gamma_{S}$. Thus, they are second class
constraints that do not generate any symmetry, but represent redundant,
non-physical quantities. They can be eliminated by defining the reduced phase
space $\Gamma^{\ast}$ with the Poisson brackets replaced by the Dirac
brackets,
\begin{align}
\left\{  A,B^{\prime}\right\}  ^{\ast}  &  =\left\{  A,B^{\prime}\right\}
+\int dy\,\left[  \rule{0pt}{15pt}\left\{  A,\lambda_{a}^{i}(y)\right\}
\left\{  p_{i}^{a}(y),B^{\prime}\right\}  -\left\{  A,p_{i}^{a}(y)\right\}
\left\{  \lambda_{a}^{i}(y),B^{\prime}\right\}  \right. \nonumber\\
&  +\left.  \frac{1}{2}\left\{  A,\lambda_{a}^{ij}(y)\right\}  \left\{
p_{ij}^{a}(y),B^{\prime}\right\}  -\frac{1}{2}\left\{  A,p_{ij}^{a}%
(y)\right\}  \left\{  \lambda_{a}^{ij}(y),B^{\prime}\right\}  \right]  \,.
\label{Dirack_brackets_}%
\end{align}
It is straightforward to check (and it is a general property of the Dirac
brackets) that the use of $\left\{  ,\right\}  ^{\ast}$ turns the weak
equality into the strong equality on $\Gamma^{\ast},$%
\begin{align}
\lambda_{a}^{i}  &  =0\,,\quad p_{i}^{a}=0\,,\qquad\text{on }\Gamma^{\ast
}\,,\nonumber\\
\lambda_{a}^{ij}  &  =0\,,\quad p_{ij}^{a}=0\,,\qquad\text{on }\Gamma^{\ast
}\,, \label{p_lambda reduced}%
\end{align}
because $\left\{  \lambda_{a}^{i},p_{j}^{\prime b}\right\}  ^{\ast}=0$, and
$\left\{  \lambda_{a}^{ij},p_{kl}^{\prime b}\right\}  ^{\ast}=0\,$\ on
$\Gamma^{\ast}$. The remaining generalized coordinates of the space
$\Gamma^{\ast}$ are $(e_{\mu}^{a},\omega_{\mu}^{ab},\pi_{a}^{\mu},\pi
_{ab}^{\mu})$ and their Dirac brackets remain unmodified (they are equal to
the Poisson brackets). From now on we drop the star from the Dirac brackets.

Let us analyze the consistency condition of $\mathcal{S}_{a}$. Using $\dot
{R}_{ij}^{bc}=D_{i}u_{j}^{bc}-D_{j}u_{i}^{bc}$ and $\dot{e}_{i}^{a}=u_{i}%
^{a},$ we get%
\begin{equation}
\mathcal{\dot{S}}_{a}=\epsilon_{abcde}\epsilon^{ijkl}\left[  20\alpha
_{0}\,D_{i}e_{t}^{b}e_{j}^{c}e_{k}^{d}e_{l}^{e}+\alpha_{2}\,D_{i}U_{j}%
^{bc}R_{kl}^{de}+\omega_{tf}^{\ \ b}\left(  20\alpha_{0}\,e_{i}^{f}e_{j}%
^{c}e_{k}^{d}e_{l}^{e}+\alpha_{2}R_{ij}^{fc}R_{kl}^{de}\right)  \right]  \,,
\end{equation}
where we denoted%
\begin{equation}
U_{i}^{ab}=u_{i}^{ab}-D_{i}\omega_{t}^{ab}, \label{U}%
\end{equation}
and used that $\left[  D_{i},D_{j}\right]  \omega_{t}^{bc}=R_{ij}^{bf}%
\omega_{tf}^{\ \ c}-R_{ij}^{cf}\omega_{tf}^{\ \ b}$. It can be recognized from
the Lagrangian formalism that $U_{i}^{ab}\equiv R_{ti}^{ab}$, because the
Hamiltonian prescription treats all time derivatives as new functions. Next,
we use a combinatorial identity, valid for any completely antisymmetric tensor
$\Sigma^{cdef}$,%
\[
0=D_{t}\epsilon_{acdef}\,\Sigma^{cdef}=\left(  \epsilon_{bcdef}\,\omega
_{ta}^{\ \ b}+2\epsilon_{abdef}\,\omega_{tc}^{\ \ b}+2\epsilon_{acdbf}%
\,\omega_{te}^{\ \ b}\right)  \Sigma^{cdef}\,.
\]
For a particular choice of $\Sigma^{fcde}=20\alpha_{0}\,e_{i}^{f}e_{j}%
^{c}e_{k}^{d}e_{l}^{e}+\alpha_{2}R_{ij}^{fc}R_{kl}^{de}, $ we obtain that
$\mathcal{S}_{a}$ does not leave the surface $\Gamma_{S}$ during its
evolution,%
\begin{equation}
\mathcal{\dot{S}}_{a}=-D_{i}\chi_{a}^{i}-\omega_{ta}^{\ \ b}\mathcal{S}%
_{b}\approx0\,.
\end{equation}
Furthermore, we also have to require the same for the torsion tensor,%
\begin{equation}
\dot{T}_{ij}^{a}=D_{i}u_{j}^{a}-D_{j}u_{i}^{a}+u_{i}^{ab}e_{bj}-u_{j}%
^{ab}e_{bi}\approx0\,.
\end{equation}
With the help of Eq.~(\ref{u_i_a}), we rewrite the last equation as%
\begin{equation}
0\approx\dot{T}_{ij}^{a}\approx R_{ij}^{ab}e_{bt}+U_{\ ji}^{a}-U_{\ ij}^{a}\,.
\end{equation}
Here the vielbein projects the Lorentz indices to the spacetime ones,
$U_{\ ji}^{a}=U_{i}^{ab}e_{bj}$. The above equation gives 30 algebraic
equations in 40 unknown functions $U_{\ ij}^{a}$, which can be decomposed into
$16+24$ components ($U_{\ ij}^{0},U_{\ ij}^{\bar{a}})$. The final solution is%
\begin{equation}
U_{\ [ij]}^{a}=\frac{1}{2}\,R_{ij}^{ab}e_{bt}=\frac{1}{2}\,R_{\ tij}^{a}%
\quad\Rightarrow\quad U_{\ [ij]}^{\mu}=\left(  0,\frac{1}{2}\,R_{\ tij}%
^{k}\right)  \,. \label{T dot}%
\end{equation}
In that way, the $6+4$ coefficients $\left(  U_{\ [ij]}^{0},U_{\ [ij]}%
^{\bar{a}}\right)  $\ become completely determined by the consistency of
$\dot{T}_{ij}^{0}$ and $\dot{T}_{[ki]j}$. Since $U_{\ ij}^{a}=R_{\ itj}^{a}$,
the above relation just represents the first Bianchi identity for the
components ($tij$) rederived in the Hamiltonian way.

The 20 components of $\dot{T}_{ij}^{a}$ that do not solve the corresponding
multipliers are\ exactly the ones symmetric in first two indices,%
\begin{align}
\dot{T}_{\left(  ki\right)  j}  &  \approx\frac{1}{2}\left(  e_{ak}\dot
{T}_{ij}^{a}+e_{ai}\dot{T}_{kj}^{a}\right) \nonumber\\
&  \approx\frac{1}{2}\,R_{ktij}+U_{k[ji]}+\frac{1}{2}\,R_{itkj}+U_{i[jk]}=0\,,
\end{align}
that vanish due to the known $U_{i[jk]}$. Thus, these components do not lead
to new conditions. We conclude that 30 equations $\dot{T}_{ij}^{a}=0$ solve
only 10 antisymmetric components $u_{i}^{ab}$ and remaining 20 equations do
not give anything new -- they are automatically satisfied.

Thanks to the relation (\ref{T dot}) and because the curvature $R_{ij}^{ab}$
satisfies the First Bianchi identity, we can collect all First Bianchi
identities in a covariant way, $\mathcal{B}_{\mu\nu\alpha\beta}=R_{\mu
(\nu\alpha\beta)}=0$, where the components of the tensor $\mathcal{B}$\ are%
\begin{align}
0  &  =\mathcal{B}_{ijk}^{a}\equiv R_{(ijk)}^{a}\,,\nonumber\\
0  &  =\mathcal{B}_{tij}^{a}\equiv e_{bt}R_{ij}^{ab}-2U_{[ij]}^{a}\,.
\end{align}
This has an important consequence on the number of \textit{linearly
independent} multipliers $U_{i}^{ab}$. Namely, we can prove that
\begin{equation}
R_{\mu\nu\alpha\beta}-R_{\alpha\beta\mu\nu}=\frac{1}{2}\,\left(
\mathcal{B}_{\mu\nu\alpha\beta}+\mathcal{B}_{\beta\mu\nu\alpha}-\mathcal{B}%
_{\alpha\beta\mu\nu}-\mathcal{B}_{\nu\alpha\beta\mu}\right)  =0\,,
\end{equation}
so the Riemann curvature is symmetric, $R_{\mu\nu\alpha\beta}=R_{\alpha
\beta\mu\nu}$, or%
\begin{equation}
R_{titj}=R_{tjti}\,,\qquad R_{tijk}=R_{jkti}\,,\qquad R_{ijkl}=R_{klij}\,.
\label{R_R_R}%
\end{equation}
The last relation in (\ref{R_R_R}) does not give any further information
because it is just the Bianchi identity on $\Sigma$. The first one, instead,
shows that not all coefficients $U_{\mu ij}=e_{a\mu}U_{ij}^{a}$, are
independent because $U_{tij}$ are symmetric, $U_{tij}=U_{tji}\,=e_{at}%
e_{bi}\,R_{tj}^{ab}$. The second condition in (\ref{R_R_R}) is equivalent to%
\begin{equation}
U_{jki}=e_{at}e_{bi}\,R_{jk}^{ab}\,, \label{u's}%
\end{equation}
in a way consistent with (\ref{T dot}). The only remaining unknown multipliers
are 10 symmetric components $U_{t(ij)}$, leading to the final expression for
$U_{i}^{ab}$ as%
\begin{equation}
U_{i}^{ab}=U_{\mu\nu i}e^{a\mu}e^{b\nu}=U_{t(ij)}\left(  e^{ta}e^{jb}%
-e^{tb}e^{ja}\right)  +e_{ct}e_{di}\,R_{jk}^{cd}e^{ja}e^{kb},\qquad
U_{tij}=U_{tji}\,. \label{Uab}%
\end{equation}

From the point of view of the irreducible components of $U_{i}^{ab}$, we can
see the 10 components of $U_{t\left(  ij\right)  }$ as the only unsolved part
in the table below,%
\[
\fbox{$%
\begin{array}
[c]{rcccccc}%
\text{Multiplier }U_{\mu ij}:\quad & U_{t[ij]}\,, & U_{tk}^{\ \ k}\,, &
\left.  ^{S}U\right.  _{tij}\,, & \left.  ^{A}U\right.  _{kij}\,, & \left.
^{S}U\right.  _{kij}\,, & \left.  ^{T}U\right.  _{kij}\,,\\
40\text{ components}:\quad & 6 & 1 & 9 & 4 & 4 & 16\\
\text{Solved by}:\quad & \dot{T}_{ij}^{0} & \text{arbitrary} &
\text{arbitrary} & \dot{T}_{[ijk]} & \text{Bianchi} & \text{Bianchi.}%
\end{array}
$}%
\]
As it is well-known, the irreducible components of the rank 2 tensor $U_{tij}
$ are: its antisymmetric part $U_{t\left[  ij\right]  }$, the trace
$U_{tk}^{\ \ k}$ and the symmetric traceless component $\left.  ^{S}U\right.
_{tij}=U_{t(ij)}-\frac{1}{4}\,g_{ij}U_{tk}^{\ \ k}$. On the other hand, the
irreducible components of the rank 3 tensor $U_{i(jk)}$ are: its vectorial
component (trace) $U_{i}\equiv U_{ij}^{\ \ j}$, also written as $\left.
^{S}U\right.  _{ijk}=g_{ij}\,U_{k}+g_{ik}\,U_{j}$, the axial-vector component
$\left.  ^{A}U\right.  _{ijk}=U_{[ijk]}$ and the tensorial one $\left.
^{T}U\right.  =U-\left.  ^{A}U\right.  -\left.  ^{S}U\right.  $.

The last equation to analyze is $\chi_{a}^{i}\approx0$. It can be combined
together with $\mathcal{H}_{a}\approx0$ into\textbf{\ }%
\begin{equation}
\chi_{a}^{\lambda}=(\mathcal{H}_{a},\chi_{a}^{i})=\epsilon_{abcde}%
\epsilon^{\lambda\mu\nu\alpha\beta}\left(  5\alpha_{0}\,e_{\mu}^{b}e_{\nu}%
^{c}e_{\alpha}^{d}e_{\beta}^{e}+\frac{\alpha_{2}}{4}\,R_{\mu\nu}^{bc}%
R_{\alpha\beta}^{de}\right)  \approx0\ ,
\end{equation}
in which we recognize the generalized Einstein equations with cosmological
constant (\ref{PL_metric_fe}). In contrast to the Einstein-Hilbert case, the
multipliers in Eq.~(\ref{v_inconclusive}) cannot be fully solved\ because they
are non-linear in the fields, causing ambiguities. In fact, if we write it as%
\begin{equation}
2\alpha_{2}\,\epsilon_{abcde}\epsilon^{ijkl}R_{kl}^{de}U_{j}^{bc}%
=-40\alpha_{0}\,\epsilon_{abcde}\epsilon^{ijkl}e_{t}^{b}e_{j}^{c}e_{k}%
^{d}e_{l}^{e}\,,\label{R_U}%
\end{equation}
then the rank of the matrix $\Omega_{abc}^{ij}=-2\alpha_{2}\epsilon
_{abcde}\epsilon^{ijkl}R_{kl}^{de}$ explicitly depends on a considered
background. More concretely, replacing the solution for the multiplier
(\ref{Uab}) in (\ref{R_U}), we obtain a set of algebraic equations
\begin{equation}
M_{a}^{i(jm)}U_{tjm}=A_{a}^{i}\,,\quad\text{or}\quad MU=A\,,\label{MU_AB}%
\end{equation}
where the matrix of the system is obtained by symmetrization of%
\begin{equation}
M_{\mu}^{ijm}=-\Omega_{abc}^{ij}\,e_{\mu}^{a}e^{tb}e^{mc}=\frac{2\alpha_{2}%
}{\left\vert e\right\vert }\,g_{\mu n}\epsilon^{mnpq}\epsilon^{ijkl}%
R_{pqkl}\,.\label{M}%
\end{equation}
The non-homogeneous part of the system is%
\begin{equation}
A_{\mu}^{i}=\left\vert e\right\vert \left(  120\alpha_{0}\,\delta_{\mu}%
^{i}-\alpha_{2}\,\epsilon_{\mu mn\nu\lambda}\epsilon^{ijkl}R_{\ \ kl}%
^{\nu\lambda}R_{tj}^{\ \ mn}\right)  \,.
\end{equation}
In the context of the equation (\ref{MU_AB}), $M$ is the $20\times10$ matrix
that acts on the 10-component column $U$. When the rank of $M$ is maximal,
that is $10$, then all components of $U$ can be determined. This is the case
of the AdS space, as we will show below. A situation is completely different
in the flat space, where the equation becomes homogeneous ($A=0$) and the rank
of\ $M=0$ is zero. In that case, all $10$ components of the vector $U_{t(jm)}$
remain arbitrary. In EH case this matrix always has maximal rank because it
does not depend on the curvature. In higher-dimensional PL gravity, $M$ is
again polynomial in the curvature and may have different ranks. It is a
generic feature of the Lovelock gravity, and it has already been noted for the
Lovelock-Chern-Simons case \cite{Banados-Garay-HenneauxD}.

We are interested in the $\Lambda\neq0$ backgrounds, where there are black
hole solutions. We restrict our theory to the part of the phase space where
the rectangular $20\times10$ matrix $M_{a}^{i(jm)}(x)$ has maximal rank, $10
$, for all $x$. In that case, there exists only the \textit{left} inverse of
$M$, which is the $10\times20$ matrix $\Delta_{(kl)i}^{a}(x)$ of the rank $10$
defined by%
\begin{equation}
\frac{1}{2}\,\Delta_{(kl)i}^{a}M_{a}^{i(jm)}=\delta_{k}^{j}\delta_{l}%
^{m}+\delta_{l}^{j}\delta_{k}^{m}\,.
\end{equation}
The matrix $\Delta$ depends on $e_{\mu}^{a}$ and $\omega_{\mu}^{ab}$. Then the
equation (\ref{MU_AB}) can be solved and the multipliers are%
\begin{equation}
U_{tij}=\Delta_{(ij)k}^{a}A_{a}^{k}\text{\thinspace}. \label{Usym}%
\end{equation}

To show that the chosen subspace contains a non-empty set of solutions, we
consider the AdS background
\begin{equation}
\bar{R}_{\mu\nu\alpha\beta}=-\frac{1}{\ell^{2}}\,\left(  \bar{g}_{\mu\alpha
}\bar{g}_{\nu\beta}-\bar{g}_{\nu\alpha}\bar{g}_{\mu\beta}\right)  \ ,
\label{AdS_back}%
\end{equation}
with $\alpha_{0}=\frac{1}{5\ell^{4}}$ and $\alpha_{2}=-1$. Then
\begin{equation}
\bar{M}_{\mu}^{i(jm)}= \frac{4}{\ell^{2}\left\vert \bar{e}\right\vert
}\,^{\left(  4\right)  }\bar{g}\ \bar{g}_{\mu n}\left(  \bar{g}^{im}\bar
{g}^{jn}+\bar{g}^{ij}\bar{g}^{nm}-2\bar{g}^{in}\bar{g}^{jm}\right)  \ ,
\label{M_def}%
\end{equation}
where $^{\left(  4\right)  }\bar{g}=\det[\bar{g}_{kl}]$ is the determinant of
the spatial background induced metric $^{\left(  4\right)  }\bar{g}_{^{ij}%
}=\bar{g}_{^{ij}}$ and its inverse is $^{\left(  4\right)  }g^{ij}%
=g^{ij}-\frac{g^{ti}g^{tj}}{g^{tt}}$.

Now we linearize Eq.~(\ref{MU_AB}) around this background, i.e., $(\bar
{M}+\delta M)(\bar{U}+V)=\bar{A}+\delta A$, where $\delta U_{t(jm)}=V_{jm}$.
We multiply the zero order, $\bar{M}\bar{U}=\bar{A}$, by $\bar{e}_{k}^{a}$ and
obtain
\begin{equation}
\bar{U}_{tij}=\frac{5\alpha_{0}\ell^{2}}{\alpha_{2}}\frac{\bar{g}\bar{g}_{ij}%
}{^{\left(  4\right)  }\bar{g}}=-\frac{1}{\ell^{2}}\,\bar{g}_{tt}\bar{g}%
_{ij}\,, \label{UAdS}%
\end{equation}
where we replaced the values of the constants $\alpha_{k}$. We also used the
identity $\left\vert \bar{e}\right\vert ^{2}=-\bar{g}=-\bar{g}_{tt}\left.
^{\left(  4\right)  }\bar{g}\right.  $.

Projecting $\bar{M}\bar{U}=\bar{A}$ by $\bar{e}_{t}^{a}$, we find that
(\ref{UAdS}) is satisfied. One can obtain the same result from the definition
$\bar{U}_{tij}=\bar{R}_{titj}$ coming from Eqs.~(\ref{U}) and (\ref{AdS_back}).

The linear order equation, $\bar{M}V+\delta M\bar{U}=\delta A$, projected by
$\bar{e}_{\mu}^{a}$ reads
\begin{equation}
\bar{M}_{\mu}^{i(jm)}V_{tjm}=C_{\mu}^{i}\,,
\end{equation}
where we defined $C_{\mu}^{i}=-\delta M_{\mu}^{i(jm)}\bar{U}_{tjm}+e_{\mu}^{a}\delta A_{a}^{i}$. After replacing the matrix $M$, see Eq.~(\ref{M_def}),
we find
\begin{equation}
V_{t\ \ \mu}^{\ i}-\delta_{\mu}^{i}V_{t\ \ j}^{\ j}= \frac{\left\vert
e\right\vert \ell^{2}}{2\left.  ^{\left(  4\right)  }\bar{g}\right.  }
\,C_{\mu}^{i}\,.
\end{equation}
In that way, all $10$ symmetric multipliers $U=\bar{U}+V$ are uniquely solved
in the AdS background with
\begin{equation}
V_{t\ \ j}^{\ i}= \frac{\left\vert e\right\vert \ell^{2}}{2\left.  ^{\left(
4\right)  }\bar{g}\right.  } \,\left(  C_{j}^{i}-\frac{1}{3}\,\delta_{j}%
^{i}C_{k}^{k}\right)  \,.
\end{equation}
It is straightforward to check that the remaining equations, $\bar{M}%
_{t}^{i(jm)}V_{tjm}=C_{t}^{i}$, are automatically satisfied. Therefore, we
explicitly found the matrix $\bar{\Delta}_{(kl)i}^{a}$ in the AdS background.

It is easy to prove in a similar way that the static black holes also belong
to the chosen region of the phase space where the left inverse $\bar{\Delta
}_{(kl)i}^{a}$ exists. Namely, the same as for the AdS space, the black hole
curvature $R_{\alpha\beta}^{\mu\nu}$ has each component proportional to
$\delta_{\alpha\beta}^{\mu\nu}$, with different factors. An explicit check
confirms that $M$ has maximal rank for static Pure Gauss-Bonnet black hole.

With all constraints identified and the Hamiltonian multipliers solved, we can
obtain the information about the degrees of freedom and local symmetries in
the theory in a particular class of backgrounds, where $M$ has maximal rank during the whole evolution of the fields.

\section{Degrees of freedom and symmetries}

Next step in the Hamiltonian analysis is to separate first and second class
constraints. The first class constraints generate local symmetries and the
second class constraints eliminate non-physical fields not related to the
symmetries. If there are $N_{1}$ first and $N_{2}$ second class constraints in
the phase space with $N$ generalized coordinates, then a physical number of
degrees of freedom is given by the Dirac formula%
\begin{equation}
N^{\ast}=N-N_{1}-\frac{1}{2}\,N_{2}\,. \label{N*}%
\end{equation}
Thus, determination of a class of constraints is of essential importance for
identification of the physical fields living on the reduced phase space
$\Gamma^{\ast}$. Furthermore, first class constraints are related to the
existence of indefinite multipliers in a theory and their numbers should match
since each first class constraint appearing in the Hamiltonian is multiplied
by an arbitrary function. Let us recall from the previous section that the
solved multipliers are $\left\{  u_{i}^{a},U_{i}^{ab},v_{a}^{i}=0,v_{a}%
^{ij}=0\right\}  $, and the unsolved ones $u_{t}^{ab}$ and $u_{t}^{a}$ are
related to the local symmetries, Lorentz transformations and diffeomorphisms.
In addition, we do not know the explicit form of all multipliers $U_{i}^{ab}$,
because $U_{t(ij)}$ depends on the background. It is then expected that we
will not be able to obtain a closed, background-independent form of all generators.

To find first class constrains, it is helpful to write the total Hamiltonian
density $\mathcal{H}_{T}$ with solved multipliers because it is known that this is first class quantity
(it commutes with all constraints) and, therefore, only first class constraints will naturally appear there as a combinations
of other constraints. Thus,  replacing the solutions (\ref{u_i_a}), (\ref{v_inconclusive}) and (\ref{U}) in $\mathcal{H}_{T}$, we obtain
the Hamiltonian density
\begin{equation}
\mathcal{H}=-\frac{1}{2}\,\omega_{t}^{ab}\,J_{ab}-e_{t}^{a}\,J_{a}
+u_{t}^{a}\pi_{a}^{t}+\frac{1}{2}\,u_{t}^{ab}\pi_{ab}^{t}+\frac{1}{2}
\,U_{i}^{ab}\phi_{ab}^{i}+\partial_{i}\mathcal{D}^{i}\,,\label{HE}
\end{equation}
where the constraints $(\mathcal{H}_{a},\mathcal{H}_{ab})$ are replaced by the new ones $(J_{a},J_{ab})$,
\begin{align}
J_{ab} &  =\mathcal{H}_{ab}-e_{ai}\pi_{b}^{i}+e_{bi}\pi_{a}^{i}+D_{i}\phi
_{ab}^{i}\nonumber\\
&  =-e_{ai}\pi_{b}^{i}+e_{bi}\pi_{a}^{i}+D_{i}\pi_{ab}^{i}\,,\nonumber\\
J_{a} &  =\mathcal{H}_{a}+D_{i}\pi_{a}^{i}\,.\label{JJ}
\end{align}
The total divergence $\mathcal{D}^{i}=e_{t}^{a}\pi_{a}^{i}+\frac{1}{2}\,\omega_{t}^{ab}\phi_{ab}^{i}$ can be neglected, as it contributes only to a
boundary term in the total Hamiltonian.

The functions $(J_{a},J_{ab})$ are not guaranteed yet to be first class because we still have to replace $U_{i}^{ab}$. But to evaluate $U\cdot\phi$, we have to
choose a particular background for $U_{t(ij)}$, so we will not write it
explicitly, as we prefer to keep the background-independent expressions. A
more detailed analysis shows that after using Eqs.~(\ref{Uab}) and
(\ref{Usym}) the multiplies can be written as
\begin{eqnarray}
\frac{1}{2}\,U_{i}^{ab}\phi _{ab}^{i} &=&\Delta
_{(ij)k}^{c}A_{c}^{k}g^{tl}e_{l}^{a}e^{jb}\phi _{ab}^{i}-e_{t}^{a}\left(
\frac{1}{2}\,R_{acjk}e^{kb}e_{i}^{c}e^{jd}\phi _{bd}^{i}-\Delta
_{(ij)k}^{c}A_{c}^{k}g^{tt}e^{jb}\phi _{ab}^{i}\right)   \nonumber \\
&=&-\frac{1}{2}\,\omega _{t}^{ab}\,\Delta J_{ab}-e_{t}^{a}\,\Delta J_{a}\,,
\end{eqnarray}
so in general this expression can affect the generators $J_{a}$ and $J_{ab}$
because $\Delta_{(ij)k}^{c}$ is a function of $e_{t}^{a}$ and $\omega_{t}
^{ab}$.
The first class generators that appear in the Hamiltonian (\ref{HE}) are
$\mathcal{J}_{ab}=J_{ab}+\Delta J_{ab}$ and $\mathcal{J}_{a}=J_{a}+\Delta J_{a}$.
Note that these corrections contain the non-linear $R^{2}$ terms and
the background-dependent $\Delta(e,\omega)$. This is similar to what happens
in the $R+T^{2}+R^{2}$ theory \cite{Blagojevic:1987qa}. Because of the
complexity of the problem, in the next step we will not account for the
$U\cdot\phi$ term.

The temporal components of the fields, $\omega_{t}^{ab}$ and $e_{t}^{a}$, are
Lagrangian multipliers because they are not dynamical, and in the Hamiltonian
notation they are arbitrary functions multiplying the constraints. Therefore,
the Hamiltonian (\ref{HE}) can be seen as the extended Hamiltonian, which
contains constraints of all generations, both primary and secondary.
Furthermore, since only first class constraints are associated to indefinite
multipliers, we can identify them as%
\[
\text{First class constraints}:\qquad \mathcal{J}_{a},\;\mathcal{J}_{ab},\;\pi_{a}^{t},\;\pi_{ab}^{t}\,,
\]
and there are $N_{1}=(5+10)\times2=30$ of them.
With respect to the second class constraints, from (\ref{Secondary}) we know that $T_{ij}^{a}\approx0$ is satisfied,
but some components of the torsion tensor are first class and some are second
class constraints. They cannot be separated explicitly. For example, $10$
functions $\mathcal{H}_{ab}$ are linear combinations of $T_{ij}^{a}$.
This means, in order to define $\mathcal{J}_{ab}$ in terms of $\mathcal{H}_{ab}$,
we had to change a basis of the constraints. In doing so, it is important that the
\textit{regularity conditions} are satisfied, ensuring that all constraints are linearly independent
on the phase space because they have the maximal rank of the Jacobian with respect to the phase space variables.
In our case, we replaced the initial set of 30 constraints $T_{ij}^{a}$ by a new one
$(\mathcal{H}_{ab},\mathcal{T}_z)$. Then, the regularity conditions require that
$\mathfrak{R}$ank$\left[\frac{\partial(\mathcal{H}_{ab},\mathcal{T}_z)}{\partial(q^M,p_N)}\right]=30$,
what means that there must be 20 second class constraints $\mathcal{T}_z$. We shall denote them by
$\mathcal{T}_z=\{\tilde{T}_{ij}^{a}\}$ regardless their tensorial properties, to remember that
they are redundant torsional components which do not generate any local symmetry. Thus, we
represented $T_{ij}^{a}$ by an equivalent set of the $10+20$ constraints
$(\mathcal{H}_{ab},\tilde{T}_{ij}^{a})$. Then we can identify the remaining
set of the constraint as
\[
\text{Second class constraints}:\qquad\tilde{T}_{ij}^{a},\,\phi_{a}^{i}%
,\,\phi_{ab}^{i}\,,
\]
and there are $N_{2}=20+20+40=80$ of them. Then the count of the degrees of
freedom is straightforward: for $N=25+50=75$ dynamical fields $(e_{\mu}%
^{a},\omega_{\mu}^{ab})$, the Dirac formula (\ref{N*}) gives the number
degrees of freedom%
\begin{equation}
N^{\ast}=5\,.
\end{equation}
This is the same number as in the five-dimensional Einstein-Hilbert theory,
and a maximal number that a PL gravity can contain. One of these degrees of
freedom is the radial one. This can be proved by performing the Hamiltonian
analysis of the action in minisuperspace approximation, which involves only
the relevant degrees of freedom, similarly as in Ref.~\cite{Giribet-Merino-Miskovic}.
Fundamental fields in this approximation are
the most general ones among $g_{\mu\nu}$ and $T_{\mu\nu\lambda}$ that have the
same isometries. The identified radial degree of freedom corresponds to the
metric component $g_{tt}=-1/g_{rr}$.

If the rank of $M$ in (\ref{MU_AB}) is smaller than maximal, then some
functions $U_{t(ij)}$ remain arbitrary, reflecting the fact that there are
more local symmetries in the theory and less degrees of freedom. In the
extreme case, when the rank of $M$ is zero, all $U_{t(ij)}$ are indefinite, so
there are $10$ additional local symmetries because second class constraints
are converted into the first class, thus $N_{1}\rightarrow N_{1}+10=40$ and
$N_{2}\rightarrow N_{2}-10=70$. This implies that in the flat background the
theory has $N^{\ast}-10+\frac{1}{2}10=0$ degrees of freedom. In general, a
number of the degrees of freedom in five-dimensional PL gravity varies in the
range%
\begin{equation}
0\leq N^{\ast}\leq5\,.
\end{equation}

Let us analyze the local symmetries and their generators. The first class
constraint $G\approx0$ acts on the fundamental field $q$ through the smeared
generator $G\left[  \lambda\right]  =\int d^{4}x\,\lambda\,G$, and the field
transforms as $\delta q=\left\{  q,G\left[  \lambda\right]  \right\}  $. In
our case, the generator for all first class constraints is%
\begin{align}
G[\Lambda,\dot{\Lambda},\epsilon,\dot{\epsilon}]  &  =\int d^{4}x\,\left[
\frac{1}{2}\,\Lambda^{ab}\left(  J_{ab}-e_{at}\pi_{b}^{t}+e_{bt}\pi_{a}%
^{t}\right)  -\frac{1}{2}\,D_{0}\Lambda^{ab}\pi_{ab}^{t}+\epsilon^{a}%
J_{a}-D_{0}\epsilon^{a}\pi_{a}^{t}\right] \nonumber\\
&  =\int d^{4}x\,\left(  \frac{1}{2}\,\Lambda^{ab}\tilde{J}_{ab}-\frac{1}%
{2}\,\dot{\Lambda}^{ab}\pi_{ab}^{t}+\epsilon^{a}\tilde{J}_{a}-\dot{\epsilon
}^{a}\pi_{a}^{t}\right)  \,,
\end{align}
where we redefined $J_{ab}\rightarrow J_{ab}-e_{at}\pi_{b}^{t}+e_{bt}\pi
_{a}^{t}$ and $J_{a}\rightarrow J_{a}+\omega_{ta}^{\ \ \ b}\pi_{b}^{t}$ in
order to covariantize the $e\cdot\pi$ term. It is equivalent to redefinition
of the multipliers, so that%
\begin{align}
\tilde{J}_{ab}  &  =J_{ab}-e_{at}\pi_{b}^{t}+e_{bt}\pi_{a}^{t}+\omega
_{ta}^{\ \ \ c}\pi_{cb}^{t}-\omega_{tb}^{c}\pi_{ac}^{t}\,,\nonumber\\
\tilde{J}_{a}  &  =J_{a}+\omega_{ta}^{\ \ \ b}\pi_{b}^{t}\,.
\end{align}
The parameters $\dot{\Lambda}^{ab}$ and $\dot{\epsilon}^{a}$ are required by
the Castellani's construction of the generators \cite{Castellani:1981us} (for an alternative method, see
\cite{Banerjee:1999yc,Banerjee:2011cu}), to
replace the independent parameters by the first class constraints
$\pi_{ab}^{t}$ and $\pi_{a}^{t}$. A reason for this is that in the Hamiltonian
formalism all PB are taken at the same time and the time derivatives of
parameters are treated as the new, independent functions, for example
$D_{0}\Lambda^{ab}$ is linearly independent of $\Lambda^{ab}$. In addition,
the Castellani method gives a procedure to determine these parameters in a way
that recovers covariance of the Lagrangian theory. Direct calculation shows
that, up to the background-dependent term $U\cdot\phi$, the given generators
indeed satisfy the Castellani's conditions.

The gauge transformations generated by $G[\Lambda,\dot{\Lambda},\epsilon
,\dot{\epsilon}]$ have the form%
\begin{align}
\delta e_{\mu}^{a}  &  =\Lambda^{ab}e_{b\mu}-D_{\mu}\epsilon^{a}\,,\nonumber\\
\delta\omega_{\mu}^{ab}  &  =-D_{\mu}\Lambda^{ab}\,.
\end{align}
The $\Lambda^{ab}(x)$ is recognized as a Lorentz gauge parameter. The local
transformations with the parameter $\epsilon^{a}(x)$ are related to the
diffeomorphisms on-shell and their explicit form cannot be written because it
depends on the background.

The non-vanishing brackets between the constraints $\{\tilde{J}_{ab},\tilde
{J}_{a},\pi_{a}^{t},\pi_{ab}^{t},T_{ij}^{a},\,\phi_{a}^{i},\,\phi_{ab}^{i}\}$
contain the Lorentz algebra%
\begin{equation}
\{\tilde{J}_{ab},\tilde{J}_{cd}^{\prime}\}=\left(  \eta_{ad}\tilde{J}%
_{bc}+\eta_{bc}\tilde{J}_{ad}-\eta_{ac}\tilde{J}_{bd}-\eta_{bd}\tilde{J}%
_{ac}\right)  \delta\,,
\end{equation}
where the brackets with weakly $\tilde{J}_{ab}$ vanish with all other
constraints, so they are explicitly first class,
\begin{align}
\{\tilde{J}_{ab},\pi_{c}^{\prime t}\} &  =\left(  \eta_{bc}\pi_{a}^{t}%
-\eta_{ac}\pi_{b}^{t}\right)  \delta\,,\nonumber\\
\{\tilde{J}_{ab},\pi_{cd}^{\prime t}\} &  =\left(  \eta_{ad}\pi_{bc}^{t}%
+\eta_{bc}\pi_{ad}^{t}-\eta_{ac}\pi_{bd}^{t}-\eta_{bd}\pi_{ac}^{t}\right)
\delta\ ,\nonumber\\
\{\tilde{J}_{ab},\tilde{J}_{c}^{\prime}\} &  =\left(  \eta_{bc}\tilde{J}%
_{a}-\eta_{ac}\tilde{J}_{b}\right)  \delta\,,\nonumber\\
\{\tilde{J}_{ab},\phi_{c}^{\prime i}\} &  =\left(  \eta_{bc}\phi_{a}^{i}%
-\eta_{ac}\phi_{b}^{i}\right)  \delta\,,\nonumber\\
\{\tilde{J}_{ab},\phi_{cd}^{\prime i}\} &  =\left(  \eta_{ad}\phi_{bc}%
^{i}+\eta_{bc}\phi_{ad}^{i}-\eta_{ac}\phi_{bd}^{i}-\eta_{bd}\phi_{ac}%
^{i}\right)  \delta\ ,\nonumber\\
\{\tilde{J}_{ab},T_{ij}^{\prime c}\} &  =\left(  \delta_{b}^{c}T_{a\,ij}
-\delta_{a}^{c}T_{b\,ij}\right)  \delta\ .
\end{align}
For completeness, we also list the other non-vanishing brackets among the
constraints,
\[
\{\tilde{J}_{a},\tilde{J}_{b}^{\prime}\}=-\frac{15\alpha_{0}}{4\alpha_{2}}\,\Omega_{abij}^{ij}\,\delta\,,
\]
where $\Omega_{abij}^{ij}=\Omega_{abcd}^{ij}\,e_{i}^{c}e_{j}^{d}$ and, with introduced  $K_{abc}^{i}=4\alpha_{2}\epsilon^{ijkl}\epsilon_{abcde}\,\omega_{j\ f}^{d}R_{kl}^{fe}+\eta_{ab}\pi_{c}^{i}-\eta_{ac}\pi_{b}^{i}$,
\begin{align}
\{\tilde{J}_{a},\pi_{bc}^{\prime t}\} &  =\left(  \eta_{ab}\pi_{c}^{t}-\eta_{ac}\pi_{b}^{t}\right)  \delta\,,\nonumber\\
\{\tilde{J}_{a},\phi_{b}^{\prime i}\} &  =-120\alpha_{0}\left\vert
e\right\vert \,\left(  e_{a}^{t}e_{b}^{i}-e_{b}^{t}e_{a}^{i}\right)
\,\delta\,,\nonumber\\
\{\tilde{J}_{a},\phi_{bc}^{\prime i}\} &  =\Omega_{abc}^{ij}\,\partial
_{j}\delta+K_{abc}^{i}\,\delta\,,\nonumber\\
\{\tilde{J}_{a},T_{ij}^{\prime b}\} &  =R_{\ aij}^{b}\delta\,,\nonumber\\
\{T_{ij}^{a},\phi_{c}^{\prime l}\} &  =-\delta_{c}^{a}\delta_{ij}%
^{lk}\,\partial_{k}\delta+\left(  \omega_{i\ c}^{a}\delta_{j}^{l}%
-\omega_{j\ c}^{a}\delta_{i}^{l}\right)  \delta\,,\nonumber\\
\{\phi_{ab}^{j},T_{kl}^{\prime c}\} &
=\left(-e_{dl}\,\delta_{ab}^{cd}\delta_{k}^{j}+e_{dk}\,\delta_{ab}^{cd}\delta_{l}^{j}\right)
\delta\,.\label{second}%
\end{align}

As already mentioned, the symplectic form in PL gravity is non-linear in the
curvature so its rank depends on the particular background. This implies that
the second class constraints cannot be in general separated from the first
class constraints. The constraints whose brackets do not vanish explicitly on
the constraint surface are the ones given by Eq.~(\ref{second}).

\section{Hamiltonian analysis of PL gravity ($d+1$)-dimensions}

In this section we only give the main results of the Hamiltonian analysis in
$d+1$ dimensions and point out the differences with respect to the
five-dimensional case. The generalized coordinates $q^{M}$, momenta $\pi_{M} $
and primary constraints have the form (\ref{generalized_coord}%
)--(\ref{Primary_const2}), where now the indices run in the wider range,
$i=1,\ldots d$ and $a=0,\ldots d$. The symplectic matrix has the components%
\begin{align}
\Omega_{abcd}^{ij}  &  =4(p-1)\beta_{p}\,\epsilon^{ijki_{4}\cdots i_{d}%
}\epsilon_{abcda_{4}\cdots a_{d}}\,R_{i_{4}i_{5}}^{a_{4}a_{5}}\cdots
R_{i_{2p-2}i_{2p-1}}^{a_{2p-2}a_{2p-1}}T_{k\,i_{2p}}^{a_{2p}}e_{i_{2p+1}%
}^{a_{2p+1}}\cdots e_{i_{d}}^{a_{d}}\,,\nonumber\\
\Omega_{abc}^{ij}  &  =\beta_{p}\,\epsilon^{iji_{3}\cdots i_{d}}%
\epsilon_{abca_{3}\cdots a_{d}}\,R_{i_{3}i_{4}}^{a_{3}a_{4}}\cdots
R_{i_{2p-1}i_{2p}}^{a_{2p-1}a_{2p}}e_{i_{2p+1}}^{a_{2p+1}}\cdots e_{i_{d}%
}^{a_{d}}\,, \label{Omega_abc}%
\end{align}
where $\beta_{p}=-2^{2-p}(d+1-2p)\,p\,\alpha_{p}$ is a real constant. The
matrix $\Omega_{abcd}^{ij}$ is identically zero only in the Einstein-Hilbert
gravity ($p=1$). In general ($p>1$), the matrix $\Omega_{abcd}^{ij}$ only
weakly vanishes for the PL gravity (\ref{PL_with_multiplier}). The phase space
functions $\mathcal{L}_{ab}^{i}$, $\mathcal{S}_{a}\,$and $\mathcal{S}_{ab}$ in
higher dimensions become%
\begin{align}
\mathcal{L}_{ab}^{i}  &  =-\frac{1}{d+1-2p}\,\Omega_{abc}^{ij}e_{j}%
^{c}\ ,\nonumber\\
\mathcal{S}_{a}  &  =\mathcal{H}_{a}+D_{i}\lambda_{a}^{i}\,,\nonumber\\
\mathcal{S}_{ab}  &  =\mathcal{H}_{ab}+e_{bi}\lambda_{a}^{i}-e_{ai}\lambda
_{b}^{i}\,,
\end{align}
where%
\begin{align}
\mathcal{H}_{ab}  &  =-\frac{1}{2}\,\Omega_{abc}^{ij}T_{ij}^{c}\,,\nonumber\\
\mathcal{H}_{a}  &  =(d+1)\,\alpha_{0}\,\epsilon_{aa_{1}\cdots a_{d}}%
\epsilon^{i_{1}\cdots i_{d}}e_{i_{1}}^{a_{1}}\cdots e_{i_{d}}^{a_{d}}-\frac
{1}{4p}\,\Omega_{abc}^{ij}R_{ij}^{bc}\,.
\end{align}
Using the Hamiltonian (\ref{Canonical_5D})--(\ref{Total_5D}), we find the
following secondary constraints from the condition of vanishing $\dot{p}%
_{ij}^{a}$, $\dot{\phi}_{a}^{t}$ and $\dot{\phi}_{ab}^{t}$,%
\begin{align}
T_{ij}^{a}  &  \approx0\,,\qquad\quad\mathcal{S}_{a}\approx
0\,,\label{consistency_d+1_a}\\
\mathcal{S}_{ab}  &  \approx e_{bi}\lambda_{a}^{i}-e_{ai}\lambda_{b}%
^{i}\approx0\,. \label{consistency_d+1_b}%
\end{align}
The requirement of vanishing $\dot{\phi}_{a}^{i}$ and $\dot{p}_{i}^{a}$ solve
the multipliers $v_{a}^{i}$ and $u_{i}^{a}$,
\begin{align}
v_{a}^{i}  &  =e_{t}^{b}\Sigma_{ab}^{i}+\frac{1}{2}\,\Omega_{cda}^{ij}%
\,U_{j}^{cd}+\omega_{t\ a}^{b}\lambda_{b}^{i}+D_{j}\lambda_{a}^{ij}%
\approx0\,,\label{v_ia_general}\\
u_{i}^{a}  &  =D_{i}e_{t}^{a}-\omega_{t}^{ab}e_{bi}\,, \label{u_ia_general}%
\end{align}
where it was convenient to define $U_{i}^{ab}=u_{i}^{ab}-D_{i}\omega_{t}^{ab}$
and%
\begin{align}
\Sigma_{ab}^{i}  &  =\epsilon_{aba_{2}\cdots a_{d}}\epsilon^{ii_{2}\cdots
i_{d}}\left[  \rule{0pt}{14pt}-d(d+1)\,\alpha_{0}e_{i_{2}}^{a_{2}}\cdots
e_{i_{2p+1}}^{a_{2p+1}}\right. \nonumber\\
&  +\left.  \frac{d-2p}{4p}\,\beta_{p}R_{i_{2}i_{3}}^{a_{2}a_{3}}R_{i_{4}%
i_{5}}^{a_{4}a_{5}}\cdots R_{i_{2p}i_{2p+1}}^{a_{2p}a_{2p+1}}\right]
e_{i_{2p+2}}^{a_{2p+2}}\cdots e_{i_{d}}^{a_{d}}\ .
\end{align}
In odd-dimensional spaces with $d=2p$, the last line in $\Sigma_{ab}^{i}$
vanishes. This is the case of five-dimensional Pure Gauss-Bonnet gravity
analyzed in previous sections.

We ask that the constraint $\phi_{ab}^{i}\approx0$ vanishes during its time
evolution in $(d+1)$-dimensional spacetime, leading to%
\begin{align}
\dot{\phi}_{ab}^{i}  &  \approx-\frac{1}{2}\,\omega_{t}^{cd}\left(
\Omega_{cda}^{ij}e_{bj}-\Omega_{cdb}^{ij}e_{aj}+\Omega_{abc}^{ij}%
\,e_{dj}-\Omega_{ab,d}^{ij}\,e_{cj}\right) \nonumber\\
&  +e_{ta}\lambda_{b}^{i}-e_{tb}\lambda_{a}^{i}-\lambda_{a}^{ij}e_{bj}%
+\lambda_{b}^{ij}e_{aj}\,. \label{two lines}%
\end{align}
However, the first line identically vanishes due to the combinatorial
identity
\begin{equation}
\epsilon_{ba_{1}\cdots a_{d}}\,e_{aj}-\epsilon_{aa_{1}\cdots a_{d}}%
\,e_{bj}+\epsilon_{aba_{2}\cdots a_{d}}\,e_{a_{1}j}-\cdots+\left(  -1\right)
^{d+1}\epsilon_{abca_{1}\cdots a_{d-1}}\,e_{a_{d}\,j}=0\,, \label{Id W}%
\end{equation}
and the second line of (\ref{two lines}) with Eq.~(\ref{consistency_d+1_b})
can be equivalently written as $\lambda_{a}^{i}\approx0$ and $\lambda_{a}%
^{ij}\approx0$, so that we find%
\begin{equation}
\text{Secondary constraints:}\qquad\mathcal{S}_{a}\approx0\,,\quad T_{ij}%
^{a}\approx0\,,\quad\lambda_{a}^{i}\approx0\,,\quad\lambda_{a}^{ij}\approx0\,.
\end{equation}

Next, we have to require that the secondary constraints also evolve on the
constraint surface. Thus, the requirement of vanishing $\dot{\lambda}_{a}%
^{ij}$ and $\dot{\lambda}_{a}^{i}$ solve the multipliers $v_{a}^{ij}=0$ and
$v_{a}^{i}=0\,$, but because the form of $v_{a}^{i}$ is already known from
Eq.(\ref{v_ia_general}), we obtain the algebraic equation for the multipliers
$U_{j}^{cd}$,%
\begin{equation}
0\approx\chi_{a}^{i}=e_{t}^{b}\Sigma_{ab}^{i}+\frac{1}{2}\,\Omega_{cda}%
^{ij}\,U_{j}^{cd}\,. \label{X_ia_d+1_v2}%
\end{equation}
By replacing $U_{i}^{ab}=R_{ti}^{ab}$, we can prove that the above expression
combined with $\mathcal{H}_{a}$ is equivalent to the Lagrangian equations,%
\begin{align}
0  &  \approx\chi_{a}^{i}=(\mathcal{H}_{a},\chi_{a}^{i})=(d+1)\,\alpha
_{0}\,\epsilon_{aa_{1}\cdots a_{d}}\epsilon^{\lambda\mu_{1}\mu_{2}\cdots
\mu_{d}}e_{\mu_{1}}^{a_{1}}e_{\mu_{2}}^{a_{2}}\cdots e_{\mu_{d}}^{a_{d}%
}\nonumber\\
&  +\frac{\alpha_{p}}{2^{p}}\,(d+1-2p)\,\epsilon^{\lambda\mu_{1}\cdots\mu_{d}%
}\epsilon_{aa_{1}\cdots a_{d}}R_{\mu_{1}\mu_{2}}^{a_{1}a_{2}}\cdots
R_{\mu_{2p-1}\mu_{2p}}^{a_{2p-1}a_{2p}}e_{\mu_{2p+1}}^{a_{2p+1}}\cdots
e_{\mu_{d}}^{a_{d}}\,.
\end{align}

Further calculation can be simplified by observing that, as in five
dimensions, the pairs of conjugated constraints, $\left(  \lambda_{a}%
^{i},p_{j}^{b}\right)  $ and $\left(  \lambda_{a}^{ij},p_{kl}^{b}\right)  $,
are second class. It means they can be eliminated from the phase space by
defining the reduced phase space $\Gamma^{\ast}$, where the Poisson brackets
are replaced by the Dirac brackets (\ref{Dirack_brackets_}). The coordinates
of the space $\Gamma^{\ast}$ are $(e_{\mu}^{a},\omega_{\mu}^{ab},\pi_{a}^{\mu
},\pi_{ab}^{\mu})$ and their Dirac brackets are equal to the Poisson brackets.
From now on, we shall drop writing the star in the Dirac brackets, and
continue working on $\Gamma^{\ast}$.

The evolution of $\mathcal{S}_{a}$ can be obtained after the long, but
straightforward calculation, with the help of the identity $0=D\epsilon
_{aba_{1}\cdots a_{d}}$, which implies%
\begin{equation}
\omega_{ta}^{\ \ b}R_{i_{1}i_{2}}^{a_{1}a_{2}}\Omega_{ba_{1}a_{2}}^{i_{1}%
i_{2}}=-\omega_{tf}^{\ \ a_{1}}\left[  2p\,\Omega_{aa_{1}a_{2}}^{i_{1}i_{2}%
}R_{i_{1}i_{2}}^{fa_{2}}+\left(  d-2p\right)  \,\Omega_{aa_{1}a_{2}a_{3}%
}^{i_{2}i_{3}k}\,R_{i_{2}i_{3}}^{a_{2}a_{3}}e_{k}^{f}\right]  \,.
\end{equation}
Then we find that $\mathcal{S}_{a}$ never leave the constraint surface,
\begin{equation}
\mathcal{\dot{S}}_{a}\approx-D_{i}\chi_{a}^{i}-\omega_{ta}^{\ \ b}%
\mathcal{H}_{b}\approx0\,.
\end{equation}
Finally, the consistency condition of $T_{ij}^{a}$ gives $\frac{\left(
d+1\right)  d\left(  d-1\right)  }{2}$ algebraic equations for $\frac
{d^{2}\left(  d+1\right)  }{2}$ unknown functions $U_{\ ij}^{a}=U_{j}%
^{ab}e_{bi}$,%
\begin{equation}
0\approx\dot{T}_{ij}^{a}\approx R_{ij}^{ab}e_{bt}+U_{\ ji}^{a}-U_{\ ij}^{a}\,.
\end{equation}
This form is the same as in five dimensions, so we skip the detailed analysis
(\ref{T dot})--(\ref{u's}) and conclude that the antisymmetric parts
$U_{\ [ij]}^{a}$ of the multipliers are solved, $U_{t(ij)}$ remain unknown and
the other are not independent due to the Bianchi identity. The final
expression for $U_{i}^{ab}$ is given by Eq.~(\ref{Uab}). The result for the
coefficients $U$ can be summarized as
\[
\fbox{$%
\begin{array}
[c]{rccccc}%
\text{Multiplier }U_{\mu ij}:\quad & U_{t[ij]}\,, & U_{t(ij)}\,, & \left.
^{A}U\right.  _{kij}\,, & \left.  ^{S}U\right.  _{kij}\,, & \left.
^{T}U\right.  _{kij}\,,\medskip\\
\frac{d^{2}\left(  d+1\right)  }{2}\text{ components}:\quad & \frac{d\left(
d-1\right)  }{2} & \frac{d\left(  d+1\right)  }{2} & d & d & \frac{d\left(
d^{2}-d-4\right)  }{2}\medskip\\
\text{Solved by}:\quad & \dot{T}_{ij}^{0} & \text{arbitrary} & \dot{T}_{[ijk]}
& \text{Bianchi} & \text{Bianchi.}%
\end{array}
$}%
\]
Solutions of $U_{t(ij)}$ depend on the equation $\chi_{a}^{i}=0$ given by
Eq.~(\ref{X_ia_d+1_v2}), which after using (\ref{Uab}) becomes
\begin{equation}
M_{a}^{i(jm)}U_{tjm}=A_{a}^{i}\,.
\end{equation}
The tensor $M$ has the form (\ref{M}). This is the $d\left(  d+1\right)
\times\frac{d\left(  d+1\right)  }{2}$ matrix of order $p-1$ in the Riemann
tensor. The non-homogenous part of the equation, on the other hand, is%
\begin{equation}
A_{a}^{i}=\Sigma_{ab}^{i}e_{t}^{b}+\frac{1}{2}\,\Omega_{abc}^{ij}e^{kb}%
e^{lc}\,R_{tjkl}\,.
\end{equation}
Higher-order dependence of $M$ in the curvature means that its rank can change
throughout the phase space. When $\Lambda\neq0$, there is always the region of
$\Gamma^{\ast}$ where the rank of $M$ is maximal, that is $\frac{d\left(
d+1\right)  }{2}$ which enables to solve all $\frac{d\left(  d+1\right)  }{2}$
coefficients $U_{t(ij)}$. This completes the constraint analysis, which has
the same structure as in five dimensions. Arbitrary multipliers are associated
with the first class constraints, and the rest are second class constraints.

Therefore, we have $N=\frac{\left(  d+1\right)  ^{2}\left(  d+2\right)  }{2}$
fundamental fields $(e_{\mu}^{a},\omega_{\mu}^{ab})$ in the PL gravity on the
reduced space, $N_{1}=\left(  d+1\right)  \left(  d+2\right)  $ first class
constraints $(J_{a},$ $J_{ab},$ $\pi_{a}^{t},$ $\pi_{ab}^{t})$ and
$N_{2}=d^{2}\left(  d+1\right)  $ second class constraints $(\tilde{T}%
_{ij}^{a},\phi_{a}^{i},\phi_{ab}^{i})$. Therefore, the number of physical
fields in the bulk in this particular background is
\begin{equation}
N^{\ast}=\frac{\left(  d+1\right)  \left(  d-2\right)  }{2}\ .
\end{equation}
In other background we can have less degrees of freedom, so that the number of
degrees of freedom in a higher-dimensional PL gravity is $0\leq N^{\ast}%
\leq\frac{\left(  d+1\right)  \left(  d-2\right)  }{2}$. The first class
constraints and gauge generators have the same form as before, only the matrix
$M$ and the tensor $A_{\mu}^{a}$\ that appear in (\ref{MU_AB}) are of order
$p-1$ in the curvature and we will not write them explicitly -- it is
straightforward to repeat the previous calculation here.

\section{{\bf Discussion}}

We performed a Hamiltonian analysis of the Pure Lovelock (PL) gravity
in any dimension $D\geq5$. This Lovelock gravity is not a mere correction of
the Einstein-Hilbert theory because it does not even contain the linear term
in the scalar curvature. Instead, its kinetic term is described by a $p$-th
order polynomial in the Riemann tensor such that the equations of motion remain of
second order in the metric. When the cosmological constant is included, the PL
gravity has the unique dS and/or AdS vacuum.

The first order formalism was used to deal with non-linearities involved in the
theory. We ensured that space-time is Riemannian by introducing the constraint
that forced the torsion to vanish.

The detailed analysis revealed that the number of symmetries and degrees of
freedom in this theory depends on the background. In the generic case, which
include (A)dS space and spherically symmetric, static black holes, the theory
contains $D(D-3)/2$ degrees of freedom, which is the same as in General
Relativity. But in contrast to Relativity, a change of the background can
increase an amount of local symmetries in the theory and convert previously
physical fields into nonphysical ones, leading even to a topological theory
(with no degrees of freedom in the bulk). This is typical for Lovelock
theories. In the PL case, this change of degrees of freedom is kept under
control through the matrix $M$, whose rank can be between 0 and $D(D-1)/2$, what
yields between 0 and $D(D-3)/2$ degrees.

A constraint analysis probes a number of physical components of the metric field $g_{\mu\nu}$,
which is directly related to the Riemann tensor. Its relation to the PL Riemann tensor is indirect and
not anchored to any metric or connection in a straightforward way. What turns out is that the maximum possible
number of physical fields does not depend on a particular
Lovelock theory, as was pointed out earlier in Ref.~\cite{Teitelboim}. It reflects the fact that so long as
the equations of motion are second order, the metric degrees of freedom would be the same for Einstein as well as Lovelock theories.

\section{Acknowledgements}

The authors thank Max Ba\~{n}ados, Milutin Blagojevi\'{c}, Branislav Cvetkovi\'{c}\ and Jorge Zanelli
for useful discussions. ND thanks Olivera Mi\v{s}kovi\'{c} for a visit to PUCV in December 2013, which facilitated the formulation of the project. This work was supported by the Chilean FONDECYT Grants No. 3140267 and 3130445, and the DII-PUCV project No. 123.738/2015.

\appendix

\section{Conventions \ \label{Conv}}

We use the signature of the Minkowski metric $\eta_{ab}=\mathtt{diag}%
(-+++\cdots+)$.

The Levi-Civita symbols in $d+1$ and $d$ dimensions are defined by
\begin{equation}
dx^{\mu_{1}}\wedge\cdots\wedge dx^{\mu_{d+1}}=\epsilon^{\mu_{1}\cdots\mu
_{d+1}}\,d^{d+1}x\,,\qquad\epsilon^{ti_{1}i_{2}\cdots i_{d}}=\epsilon
^{i_{1}i_{2}\cdots i_{d}}\,.
\end{equation}
The generalized Kronecker delta of rank $s$ is constructed as the determinant
\begin{equation}
\delta_{\mu_{1}\cdots\mu_{s}}^{\nu_{1}\cdots\nu_{s}}=\left\vert
\begin{array}
[c]{cccc}%
\delta_{\mu_{1}}^{\nu_{1}} & \delta_{\mu_{1}}^{\nu_{2}} & \cdots & \delta
_{\mu_{1}}^{\nu_{s}}\\
\delta_{\mu_{2}}^{\nu_{1}} & \delta_{\mu_{2}}^{\nu_{2}} &  & \delta_{\mu_{2}%
}^{\nu_{s}}\\
\vdots &  & \ddots & \\
\delta_{\mu_{s}}^{\nu_{1}} & \delta_{\mu_{s}}^{\nu_{2}} & \cdots & \delta
_{\mu_{s}}^{\nu_{s}}%
\end{array}
\right\vert \,.
\end{equation}
If the range of indices is $D$, a contraction of $k\leq s$ indices in the
Kronecker delta of rank $s$ produces a delta of rank $s-k$,
\begin{equation}
\delta_{\mu_{1}\cdots\mu_{k}\cdots\mu_{s}}^{\nu_{1}\cdots\nu_{k}\cdots\nu_{s}%
}\,\delta_{\nu_{1}}^{\mu_{1}}\cdots\delta_{\nu_{k}}^{\mu_{k}}=\frac{\left(
D-s+k\right)  !}{\left(  D-s\right)  !}\,\delta_{\mu_{k+1}\cdots\mu_{s}}%
^{\nu_{k+1}\cdots\nu_{s}}\,.
\end{equation}
Other identities involving the Levi-Civita symbol and the generalized
Kronecker delta are
\begin{align}
\epsilon_{\nu_{1}...\nu_{d+1}}\epsilon^{\mu_{1}...\mu_{d+1}}  &  =-\delta
_{\nu_{1}...\nu_{d+1}}^{\mu_{1}...\mu_{d+1}}\,,\nonumber\\
\epsilon_{a_{1}\cdots a_{d+1}}e_{\mu_{1}}^{a_{1}}\cdots\ e_{\mu_{d+1}%
}^{a_{d+1}}  &  =|e|\epsilon_{\mu_{1}...\mu_{d+1}}\,,
\end{align}
where $|e|=\det[e_{\alpha}^{a}]$.

\end{document}